\documentclass[12pt,aps,prd,preprint,tightenlines,
superscriptaddress,fleqn,
nofootinbib]{revtex4-1}

\RequirePackage[colorlinks=true
,urlcolor=blue
,anchorcolor=blue
,citecolor=blue
,filecolor=blue
,linkcolor=blue
,menucolor=blue
,pagecolor=blue
,linktocpage=true
,pdfproducer=medialab
,pdfa=true
]{hyperref}

\usepackage{amsmath,amssymb,amsfonts}
\allowdisplaybreaks
\usepackage{graphicx}
\usepackage{subfigure} 
\usepackage{dcolumn}
\usepackage{hyperref}      
\usepackage{bm}
\usepackage{setspace}
\usepackage[usenames, dvipsnames]{color}
\usepackage{cancel}
\usepackage{comment}
\usepackage{mathtools}
\usepackage{bbm} 
\usepackage{fleqn} 
\usepackage{mathrsfs}	
\usepackage{mathtools}
\usepackage{bbm}
\usepackage{commath}
\usepackage{physics}
\usepackage{siunitx}
\usepackage{datetime}
\usepackage{tikz}
\usepackage{cleveref} 
\newcommand{\PRE}[1]{{#1}} 
\newcommand{\CenterObject}[1]{\ensuremath{\vcenter{\hbox{#1}}}}
\DeclareMathOperator*{\SumInt}{%
\mathchoice%
  {\ooalign{$\displaystyle\sum$\cr\hidewidth$\displaystyle\int$\hidewidth\cr}}
  {\ooalign{\raisebox{.14\height}{\scalebox{.7}{$\textstyle\sum$}}\cr\hidewidth$\textstyle\int$\hidewidth\cr}}
  {\ooalign{\raisebox{.2\height}{\scalebox{.6}{$\scriptstyle\sum$}}\cr$\scriptstyle\int$\cr}}
  {\ooalign{\raisebox{.2\height}{\scalebox{.6}{$\scriptstyle\sum$}}\cr$\scriptstyle\int$\cr}}
}
\newcommand{\I}{\ensuremath{\mathrm{i}}} 
\newcommand{\be}{\begin{equation}\begin{aligned}}
\newcommand{\ee}{\end{aligned}\end{equation}}
\newcommand{\beq}{\begin{equation}}
\newcommand{\eeq}{\end{equation}}
\newcommand{\beqa}{\begin{eqnarray}}
\newcommand{\eeqa}{\end{eqnarray}}

\newcommand{\mplanck}{M_{\text{P}}}
\newcommand{\FlavonVEV}{\ensuremath{v_S}}

\newcommand{\gev}{\si{\giga\electronvolt}}
\newcommand{\tev}{\si{\tera\electronvolt}}

\newcommand{\eg}{{\em e.g.}}
\newcommand{\ie}{{\em i.e.}}

\newcommand{\eqsref}[2]{Eqs.~(\ref{#1}) and (\ref{#2})}

\newcommand{\z}{\ensuremath{u}}

\newcommand{\UFN}{\ensuremath{\text{U}(1)_\mathrm{FN}}}

\DeclareMathOperator{\re}{Re}
\DeclareMathOperator{\im}{Im}

\DeclareMathOperator{\arctanh}{arctanh}

\newcommand{\mpl}{\ensuremath{M_\text{P}}}

\newcommand{\tewk}{\ensuremath{{T_{\rm EWPT}}}}

\newcommand{\xmax}{\ensuremath{{x_\text{max}}}}
\newcommand{\xmin}{\ensuremath{{x_\text{min}}}}
\newcommand{\barQ}{{\overline{Q}}}
\newcommand{\msq}{\ensuremath{\abs{\mathcal M}^2}}

\newcommand{\ms}{\ensuremath{m_\sigma}}

\newcommand{\tg}{\tilde{g}}
\newcommand\Tstrut{\rule{0pt}{4ex}}  
\newcommand{\widesim}[2][1.5]{
  \mathrel{\overset{#2}{\scalebox{#1}[1]{$\sim$}}}
}

\newcommand{\tdash}{\protect\raisebox{2.5pt}{\protect\rule{4pt}{0.3pt}}}

\begin{document}

\preprint{UCI-TR-2018-04}

\title{\texorpdfstring{\PRE{\vspace*{0.5cm}}}{} The Flavor of Cosmology\texorpdfstring{\vspace*{0.5cm}}{}}

\author{Benjamin Lillard}
\email{blillard@uci.edu}
\affiliation{Department of Physics and Astronomy, University of
California, Irvine, CA 92697-4575 USA}

\author{Michael Ratz}
\email{mratz@uci.edu}
\affiliation{Department of Physics and Astronomy, University of
California, Irvine, CA 92697-4575 USA}

\author{Tim M.P. Tait}
\email{ttait@uci.edu}
\affiliation{Department of Physics and Astronomy, University of
California, Irvine, CA 92697-4575 USA}
\affiliation{Institute of Physics and Astronomy, University of Amsterdam, The Netherlands}

\author{Sebastian Trojanowski}
\email{strojano@uci.edu \& Sebastian.Trojanowski@ncbj.gov.pl}
\affiliation{Department of Physics and Astronomy, University of
California, Irvine, CA 92697-4575 USA}
\affiliation{National Centre for Nuclear Research,\\Ho{\. z}a 69, 00-681 Warsaw, Poland
\vspace*{2cm}}


\begin{abstract}
\PRE{\vspace*{0.5cm}}
We discuss the cosmology of models in which the standard model Yukawa couplings 
depend on scalar field(s), often referred to as flavons. We find that
thermal corrections of the flavon potential tend to decrease the Yukawa
couplings, providing an important input to model{\tdash}building. Working in the
specific framework of Froggatt--Nielsen models, we compute the abundance of
flavons in the early universe generated both via freeze{\tdash}in and from coherent
oscillations induced by thermal corrections to their potential, and discuss
constraints on flavon models from cosmology.  We find that cosmology places
important constraints on theories containing flavons even for regions of
parameter space inaccessible to collider searches.
\end{abstract}

\maketitle
\pagenumbering{gobble}

\section{Introduction}
\label{sec:introduction}
\pagenumbering{arabic}
\setcounter{page}{2}

Despite its tremendous phenomenological success, the Standard Model of particle
physics (SM) leaves open a number of key questions.  In particular, the
surprisingly large hierarchies manifest in the masses of the fermions and their
peculiar pattern of mixing seems to hint at an underlying dynamical structure. 
Many extensions of the SM have been proposed to address this so{\tdash}called flavor
puzzle~\cite{Georgi:1972hy,Froggatt:1978nt,Kaplan:1991dc,ArkaniHamed:1999dc,Gherghetta:2000qt,Kaplan:2001ga,Grossman:1999ra,Blanke:2008zb,Casagrande:2008hr,Bauer:2009cf}.
Among them, arguably one of the most compelling proposals is the idea by
Froggatt and Nielsen \cite{Froggatt:1978nt} that the pattern of masses and
mixings reflects the charges of a spontaneously broken, family{\tdash}non{\tdash}universal 
$\UFN$ symmetry.

While Froggatt--Nielsen (FN) models can arise from a variety of different 
ultraviolet (UV) theories, they share two common ingredients:  an Abelian $\UFN$
symmetry and the presence of at least one flavon $S$ -- a scalar field whose
vacuum expectation value  (VEV) spontaneously breaks $\UFN$. The couplings of
$S$ to pairs of SM fermions are dictated by the $\UFN$ charges of the latter,
and the mass hierarchies and mixing patterns are encoded in terms of powers of
the ratio $\varepsilon$ between $\langle S\rangle$ and the UV completion scale
$\Lambda$, but are relatively insensitive to the scale of $\Lambda$ itself. 
Typically, $\varepsilon \equiv \langle S\rangle / \Lambda$ is of the order of
the Cabbibo angle, and the entries of the Yukawa couplings are (up to order one
coefficients descending from the UV theory) given by powers of $\varepsilon$. 

In this study, we examine the cosmological constraints on models containing
flavons. Our focus is the parameter region in which $\Lambda$ is large,
as suggested by constraints from flavor observables, but
the flavon excitations are comparatively light.  In this regime,
late decays of the flavon pose two
possible threats to standard cosmology: they may spoil the successful
predictions from primordial nucleosynthesis (BBN) and/or dilute the primordial
baryon asymmetry to an unacceptable level.
We show that cosmology leads to interesting bounds on the parameter space.

In fact,
though our explicit calculations are performed in the specific framework of
Froggatt--Nielsen models, the constraints we derive apply to a much
broader range of frameworks in which the Yukawa interactions
are determined by VEV's. In particular, our discussion reveals that the
so{\tdash}called moduli problem~\cite{Coughlan:1983ci,deCarlos:1993wie}
typically discussed in the context of string model building
is more severe than previously appreciated.

This paper is organized as follows. In \Cref{sec:FNmodel} we review some of the basic
properties of FN models relevant for our analysis. In
\Cref{sec:flavonproduction} we discuss various flavon production mechanisms in the early
Universe, and in \Cref{sec:cosmoconstraints}, derive cosmological bounds. 
We conclude in \Cref{sec:conclusions}. 
Some of the technical details are presented in more detail in a number of appendices.

\section{Froggatt--Nielsen}
\label{sec:FNmodel}

The low{\tdash}energy effective Lagrange density of an FN model contains a global $\UFN$
symmetry such that the Yukawa couplings for the SM  quarks are realized as 
non{\tdash}renormalizable interactions containing the appropriate power of $S$ to
maintain $\UFN$,
\begin{equation}
 \mathscr{L}_\mathrm{FN}~=~\sum_{i,j=1}^3y^u_{ij}\,
 \left(\frac{S}{\Lambda}\right)^{n_{ij}^u}\,\overline{Q}_i\,\widetilde{\Phi}\,u_j 
 + \sum_{i,j=1}^3y^d_{ij}\,\left(\frac{S}{\Lambda}\right)^{n_{ij}^d}\,
 \overline{Q}_i\,\Phi\,d_j + \text{h.c.}\;,\label{eq:L}
\end{equation}
where, $y^{u/d}_{ij}$ are dimensionless couplings presumably of order unity, 
and $\Phi$, $Q_i$, $u_i$ and $d_j$ denote the SM Higgs and quark fields. 
The integer powers $n_{ij}^{u/d}$ derive from the $\UFN$ charges
of the respective quarks. In a convention in which $S$ carries $\UFN$ charge
$Q_\mathrm{FN}(S)=-1$, one has $n_{ij}^u=-Q_\mathrm{FN}(Q_i)+Q_\mathrm{FN}(u_j)$
and $n_{ij}^d=-Q_\mathrm{FN}(Q_i)+Q_\mathrm{FN}(d_j)$, respectively. The $\UFN$
symmetry forbids direct couplings between Higgs and most of the SM quarks and
leptons. 
A non{\tdash}zero VEV for $S$ results in hierarchical Yukawa couplings given by
$\varepsilon=\FlavonVEV/(\sqrt{2}\Lambda)$ raised to the appropriate power
which, together with appropriate order one coefficients $y_{ij}^{u/d}$ reproduce
the observed fermion masses and mixings
\cite{Ibanez:1994ig,Binetruy:1994ru,Jain:1994hd,Dudas:1995yu,Nir:1995bu,Binetruy:1996xk}.
Reproducing the quark masses and mixings fix, to some extent, 
$\varepsilon$ and the $\UFN$ charges, but leave the UV scale $\Lambda$ and the flavon mass
$m_\sigma$ as free parameters.

In this work, we choose to focus exclusively on the quarks, as they typically dominate flavon phenomenology
and leave exploration of analogous constructions for the leptons to future work.
Because of the freedom in the $y^{u/d}_{ij}$ coefficients, the 
FN charges are not uniquely determined by the observed quark masses and mixings.
This leaves some model{\tdash}dependence in the flavon couplings.  In 
\Cref{sec:flavoncouplings}, we present two representative choices of parameters
based on $\varepsilon=0.23$ and
reproducing the observed quark masses and mixings, which
we use in our numerical studies.

There are a variety of UV complete models whose low energy limit is \Cref{eq:L} 
(see e.g.\ \cite{Leurer:1992wg,Leurer:1993gy,Ramond:1993kv}).  
While the precise details of the UV completion
are typically not very important for our purposes, they can serve as a guide for interesting regions of parameter space.
In most explicit models, the $\UFN$ charges of the (left{\tdash}chiral) spinors
describing SM fermions all have the same sign, and the $\UFN$ symmetry is
anomalous (however, see \cite{Chen:2008tc}). It is well
known that pseudo{\tdash}anomalous $\UFN$ symmetries can arise from string
theory~\cite{Binetruy:1994ru,Binetruy:1996xk}, with the anomalies cancelled
by the Green--Schwarz mechanism~\cite{Green:1984sg}, and the VEV of the flavon
set by a (field{\tdash}dependent) Fayet--Iliopoulos term \cite{Fischler:1981zk,Dine:1987xk}. 
In such cases, $\Lambda$ is expected to be of the order of
the Planck scale $\mplanck$. On the other hand, in string models the fields
determining the couplings, often referred to as moduli, usually have masses well
below $\Lambda$. While explicit string models tend to have richer structure
than the FN model, analogous constraints also typically apply to them.

There are also models in which the flavor scale is as
low as $\Lambda \sim$~TeV such that flavons can in principle be produced at the Large Hadron Collider
(LHC)~\cite{Tsumura:2009yf,Bauer:2016rxs}. 
For lighter masses, there are typically strong bounds from flavor changing neutral
current (FCNC) processes (for recent discussion see e.g.~\cite{Calibbi:2012at})
with typical limits of order $\sqrt{\Lambda\,m_\sigma}\gtrsim
\text{few}~\tev$~\cite{Baldes:2016gaf}. 
We will consider masses as low as $m_\sigma\gtrsim \si{10~\giga\electronvolt}$,
and $\Lambda$ in the range $\tev\lesssim\Lambda\lesssim\mplanck$.

\subsection{\boldmath$\UFN$--Breaking\unboldmath}

In order to generate the Yukawa interactions, one spontaneously breaks
the $\UFN$ symmetry by
engineering a potential for the flavon,
\begin{equation}\label{eq:VS}
\mathscr{V}_S ~=~ -\mu_S^2|S|^2+\lambda_S|S|^4 + \lambda_{S\Phi} |S|^2 |\Phi|^2
+\text{\UFN\ breaking terms}\;,
\end{equation}
such that the flavon acquires a non{\tdash}zero VEV $\FlavonVEV$,
\begin{equation}\label{eq:FlavonS}
S ~=~\frac{1}{\sqrt{2}}(\FlavonVEV + \sigma + \I\,\rho)\;.
\end{equation}
The mixed quartic $\lambda_{S\Phi}$ between the flavon and the Higgs is unavoidable, but
we will assume it is small enough that it can be safely ignored in our analysis.  If present,
it induces mixing between the flavon and the Higgs boson after both
$\UFN$ and electroweak symmetry-breaking, and contributes to the mass parameters
for both flavon and Higgs.

Soft $\UFN$ breaking terms are included to give mass to the pseudoscalar flavon
$\rho$, lifting it from the effective theory.
A very light $\rho$ would be ruled out, whereas $m_\rho \sim m_\sigma$ would
result in an order one change in the cosmological bounds we derive below.
We focus on the limit $m_\rho \gg m_\sigma$ to simplify our analysis.
Of course, invoking explicit $\UFN$ breaking raises the possibility 
of a loss of predictivity such that the explanation
of the fermion mass hierarchy and mixing structure will be spoiled.
In \Cref{sec:scalarpotential}, we construct a model in which $\UFN$ is replaced by a discrete
$\mathbbm{Z}_N$ symmetry, which allows for the $\rho$ to be heavier than $\sigma$
while maintaining the FN mechanism.

\subsection{Flavon Couplings}

Before the electroweak phase transition (EWPT), the flavon couples to pairs of
quarks and the Higgs boson via non{\tdash}renormalizable interactions obtained from
\Cref{eq:L} by expanding $S$ around its VEV,
\begin{equation}
\mathscr{L}_{\substack{\mathrm{before}\\[1pt] \mathrm{EWPT}}}~\supset~
\sum_{i,j}{\frac{g^u_{ij}}{\Lambda}\sigma\,
\overline{Q}_i\widetilde{\Phi}u_j} + \sum_{i,j}{\frac{g^d_{ij}}{\Lambda}\sigma 
\overline{Q}_i\widetilde{\Phi}d_j}\;,
\label{eq:Lbefore}
\end{equation}
where
\begin{equation}
g_{ij}^u ~\simeq~\frac{y_{ij}^u}{\sqrt{2}}\,n^u_{ij}\,
\varepsilon^{n^u_{ij}-1}
\quad\text{and}\quad
g_{ij}^d~\simeq~\frac{y_{ij}^d}{\sqrt{2}}\,n^d_{ij}\,
\varepsilon^{n^d_{ij}-1}\;.
\label{eq:gbefore}
\end{equation}

After EWPT, the Higgs can be replaced by its VEV, $v_\mathrm{EW}\simeq246\,\gev$, producing
renormalizable interactions.  In the quark mass basis $u_{\mathrm{L},\mathrm{R}}$ and
$d_{\mathrm{L},\mathrm{R}}$, the flavon couplings are given by
\begin{equation}
\mathscr{L}_{\substack{\mathrm{after}\\[1pt] \mathrm{EWPT}}}~\supset ~
\tilde{g}_{ij}^u\,\sigma\,\bar{u}_{\mathrm{L},i}\,u_{\mathrm{R},j} + \tilde{g}_{ij}^d\,\sigma\,\bar{d}_{\mathrm{L},i}\,d_{\mathrm{R},j}\;,
\label{eq:Lafter}
\end{equation}
where
\begin{equation}
\tilde{g}_{ij}^u ~\simeq~\frac{1}{2\,\varepsilon}\frac{v_\mathrm{EW}}{\Lambda}\,\left[U_{u}^\dagger\,(n^u Y_u)
W_{u}\right]_{ij}
\quad\text{and}\quad
\tilde{g}_{ij}^d ~\simeq~\frac{1}{2\,\varepsilon}\frac{v_\mathrm{EW}}{\Lambda}\,\left[U_{d}^\dagger\,(n^d Y_d)
W_{d}\right]_{ij}\;.
\label{eq:gafter}
\end{equation}
The $Y_{u/d} \equiv y^{u/d}_{ij}\,\varepsilon^{n^{u/d}_{ij}}$ are the Yukawa interactions, 
which are diagonalized as usual by unitary matrices $U_{u/d}$ and $W_{u/d}$ via the biunitary transformation
$Y_{u/d}^\mathrm{diag} = U_{u/d}^\dagger\,Y_{u/d}\,W_{u/d}$.
The object $(n Y) \equiv n_{ij}Y_{ij}$ is not generally diagonal in the mass basis, leading the flavon to mediate
tree level flavor{\tdash}changing neutral currents.

Note that all interactions between the flavon and quarks are suppressed by at least one power of
$\Lambda$, and vanish in the limit of very large flavor scale.

\section{Flavon Production in the Early Universe}
\label{sec:flavonproduction}

Flavons can be produced in the early universe through a variety of processes.
As usual, the key question is whether or not the flavon chemically equilibrates with the SM plasma
in the early Universe.  
If the Hubble expansion rate $H$ exceeds the rate of flavon production at all relevant times, $\Gamma_{\textsc{SM}\leftrightarrow \sigma} < H$,
the flavon is too weakly coupled to achieve thermal equilibrium in the early universe, and the flavon abundance will be determined by
out{\tdash}of{\tdash}equilibrium processes. 

At high temperatures (above the EWPT), the flavon's interactions with quarks are nonrenormalizable, and fall out of equilibrium
at a temperature $T_\text{dec}$, defined by
$H(T_\text{dec}) = n_X \langle \sigma v \rangle$, where the
thermally averaged cross section in the limit $T \gg \ms$ is $\langle \sigma v \rangle \sim (g^{u/d}_{ij})^2 /\Lambda^2$, 
and where $n_X \simeq n_X^\mathrm{eq}= \zeta(3) T^3 /(4
\pi^3)$ for relativistic bosons.  It is convenient to write $H=T^2/M_H$ with
$M_H=\mpl/(1.66 \sqrt{g_*^\rho}) \approx 1.4\times 10^{17}~\gev$ for the
radiation{\tdash}dominated universe. Up to $\mathcal O(1)$ factors,
\begin{equation}
T_\mathrm{dec}~\approx~\frac{10^2 \Lambda^2}{M_H \sum_{ij} \abs{g^{u/d}_{ij}}^2} \sim 10^{15}~\gev \left(\frac{\Lambda}{10^{15}\ \gev} \right)^2
\;.
\label{eq:Tfo}
\end{equation} 

Typically the highest relevant temperature correspond to the reheating temperature after inflation, $T_R$.
For $T_R < T_\text{dec}$, the flavon abundance is set by out{\tdash}of{\tdash}equilibrium
processes:
\begin{itemize}
\item For most of the parameter space, the dominant production mechanism is from model{\tdash}independent
thermal corrections to the flavon potential.
\item Flavons are also produced through the scattering of standard model particles, in parallel 
production of gravitinos~\cite{Bolz:2000fu,Pradler:2006qh,Rychkov:2007uq} and
axinos~\cite{Covi:2001nw} in supersymmetric scenarios, and more generally
with  ``freeze{\tdash}in" production~\cite{Hall:2009bx}.
One interesting subtlety arises because the flavon couplings become effectively renormalizable
after the EWPT, leading to two distinct epochs of freeze-in production.
 \item There are potentially additional model{\tdash}dependent production mechanisms.  For example, there may be direct couplings between the flavon and inflaton,
leading to flavon production when the inflaton decays.  Or the flavon could experience a brief period of equilibrium during
the reheating period, in which the temperature can attain much larger values than $T_R$, though this is typically
balanced by efficient dilution in the fast expanding Universe. 
We neglect these contributions to the yield, leading to results 
which are conservative in the sense that their inclusion could only lead to more stringent cosmological constraints.
\end{itemize}
We examine the model{\tdash}independent processes in more detail in the following subsections.

\subsection{Finite Temperature Corrections to the Scalar Potential} \label{sec:scalarF}

At temperatures $T\gtrsim \order{\ms}$, thermal corrections to the effective
flavon potential $\mathscr{V}(\sigma)$ supplement the freeze{\tdash}in contribution to
the flavon yield by inducing oscillations in $\sigma$. 
The thermal corrections to the potential of the flavon are analogous to those to
the dilaton potential \cite{Buchmuller:2003is,Buchmuller:2004xr}. This discussion applies to the
case in which the flavon is not in thermal equilibrium. For a discussion of
equilibrated flavons, see e.g.~\cite{Ema:2018abj}. The crucial point is that the
free energy receives contributions from the Yukawa couplings $y$, which we derive in 
\Cref{sec:FreeEnergyYukawa},
\begin{equation}
\mathcal{F}~=~ \gamma\, y^2\, T^4 + \alpha\, T^4 \frac{\sigma}{\Lambda} + \ldots\;,
\end{equation}
where $\gamma=5/96$. As the latter depend on the flavon, this leads to a
thermal potential of the flavon of the form
\begin{equation}\label{eq:Veff}
 \mathscr{V}_\mathrm{eff}(\sigma, T)~=~ \gamma\, T_Y\, T^4 + 
 \alpha\, T^4 \frac{\sigma}{\Lambda} + \frac{m_\sigma^2(T)}{2}\,\sigma^2+ 
  \frac{\kappa}{3!}\,  \sigma^3 + \frac{\lambda_S}{4} \, \sigma^4 +
 \ldots\;,
\end{equation}
where, $\kappa \sim \ms^2 /v_s$, $T_Y=\tr\left(Y_u^\dagger Y_u+Y_d^\dagger Y_d\right)$ and 
\begin{equation}
 \alpha~=~\gamma\,\frac{\partial T_Y}{\partial \varepsilon}~\sim~10^{-2}\;.
\end{equation}
The $\alpha$~term in \Cref{eq:Veff} drives the flavon away from its
zero{\tdash}temperature minimum towards smaller values.

Qualitatively, the flavon gets driven away from
its $T=0$ minimum until it gets stopped by the mass term or Hubble friction,
\begin{equation}\label{eq:DeltaSigma}
\Delta\sigma~\simeq~-\alpha\,\frac{T^4}{\Lambda\,m_\mathrm{eff}^2}
\qquad\text{where}~m_\mathrm{eff}^2~=~6H^2+m_\sigma^2\;.
\end{equation}
As the temperature decreases, the flavon undergoes oscillations around the $T=0$
minimum, which behave like nonrelativistic matter.

More quantitatively,
the energy density in the form of flavon oscillations follows the equation of motion,
$\ddot{\sigma} + (3H + \Gamma_\sigma) \dot\sigma + \partial \mathscr{V}_\mathrm{eff}/\partial \sigma = 0$, 
which together with $\dot T \approx - H\, T$ becomes
\begin{equation}
H^2\, T^2\, \frac{\dd^2 \sigma}{\dd T^2} - 
\Gamma_\sigma\, H\, T\,  \frac{\dd \sigma}{\dd T} 
+ \alpha\, \frac{T^4}{\Lambda} + \ms^2 \sigma 
+ \frac{1}{2} \kappa \sigma^2 + \lambda_s \sigma^3 ~\approx~ 0\;.
\end{equation}
Under the assumptions that $T \gg \Gamma_\sigma $ and $\lambda_S \ll 1$, this equation of motion can be approximated by
\begin{equation}
 H^2\, T^2\, \frac{\dd^2 \sigma}{\dd T^2} + 
 \alpha \frac{T^4}{\Lambda} + \ms^2\, \sigma ~\approx~ 0\;. \label{eq:sigmaEOM}
\end{equation}

The exact solution to \Cref{eq:sigmaEOM} is detailed in \Cref{sec:oscillations}, but it is instructive to discuss limiting cases.
At high temperatures $H \gg \ms$, the two asymptotic solutions for $\sigma(T)$
scale as $\sigma \propto T$ and $\sigma \propto \ln T$, while at low{\tdash}to{\tdash}intermediate
temperatures $\Gamma_\sigma \ll H \lesssim \ms$ the solutions in a radiation{\tdash}dominated 
universe ($H=  T^2/M_H$) approach the
form (see \Cref{fig:OscillationAmplitude} in \Cref{sec:oscillations})
\begin{equation}
 \sigma(T)~=~\sigma_0\, 
 \left( \frac{T}{T_*} \right)^{3/2} \cos \left( \frac{1}{2} 
 \frac{T_*^2}{T^2} - \varphi \right)
 \qquad\text{with}~T_*~\equiv~\sqrt{\ms\, M_H}
 \;, \label{eq:oscsigma}
\end{equation}
where $\varphi$ denotes a phase that is determined by the initial conditions.

The amplitude $\sigma_0$ is highly sensitive to the boundary conditions if $T_R \gtrsim T_*$, 
but tends to $\sigma_0\propto T_R^{5/2}$ if $T_R \ll T_*$. This
scaling with $T_R$ continues at high temperature if one assumes that the flavon
$\sigma$ begins at rest at the minimum of its \emph{finite} temperature
potential $\mathscr{V}(T=T_R)$ rather than at $\mathscr{V}(T=0)$. %
A different boundary condition $\sigma(T_R) = \dot \sigma(T_R) = 0$ gives an
amplitude proportional to $\ln (T_R/T_*)$ at high temperatures. 
We adopt a rather conservative {\em lower} bound on the flavon yield based on the 
$\sigma(T_R) = \dot \sigma(T_R) = 0$ boundary condition. 
A more general discussion of the initial conditions is provided in \Cref{sec:oscillations}. 

From the amplitude of the oscillations in \eqref{eq:oscsigma}, the yield from
flavon oscillations reads
\begin{equation}\label{eq:Yosc}
Y^\mathrm{osc} \approx \alpha^2 \frac{A_* \mpl \ms}{\Lambda^2} 
\left[\frac{M_H (T_R)}{\ms} \right]^{3/2} \times 
\begin{dcases}
\left( \frac{T_R}{T_*} \right)^{5} &~ \text{if}~T_R 
\lesssim T_*\;, \\ \Tstrut
2.1 \ln^2\left(\frac{T_R}{T_*} \right) &~\text{if}~ T_R \gg T_*
\; .
\end{dcases}
\end{equation}
Note that it is possible for the energy density in the flavon oscillations to exceed that
of the relativistic thermal bath, causing an early matter{\tdash}dominated epoch in
the evolution of the universe.
In this event the assumption $H = T^2/M_H$ ceases to apply, and the evolution of
$\sigma(T)$ is no longer described by \Cref{eq:oscsigma}. However, in the
absence of an additional entropy production from other sources, the flavon yield 
at the time of decay is still described by \Cref{eq:Yosc}. 


\subsection{High Temperature Freeze{\tdash}In Production} 
\label{sec:yUV}

At temperatures above the electroweak phase transition, flavon production occurs
via $2\rightarrow 2$ scattering through the nonrenormalizable operators, \eqsref{eq:Lbefore}{eq:gbefore}.
For every pair of quarks $Q_{i}$
and $u_{j}$, there are six processes by
which the flavon $\sigma$ are produced,
\begin{align}
\barQ_{i} + u_{j} &\rightarrow \sigma + H \;,&
\barQ_{i} + H^\dagger &\rightarrow \sigma + \bar u_{j}\;, &
H^\dagger + u_{j} &\rightarrow \sigma + Q_{i}\;,\notag\\
Q_{i} + \bar u_{j} &\rightarrow \sigma + H^\dagger ,& 
Q_{i} + H &\rightarrow \sigma +  u_{j} \;, &
H + \bar u_{j} &\rightarrow \sigma + \barQ_{i}\;.
\label{eq:UVprocess}
\end{align}
A similar set of processes exists for the down{\tdash}type quarks. Adding them together
and solving the respective Boltzmann equation one obtains the ultraviolet (UV) contribution to the total
yield of flavon $Y_\sigma = n_\sigma/s$, where $n_\sigma$ is the flavon number
density, and $s$ is the entropy density. For a given process involving up or down
quarks from the $i$th and $j$th generation, it reads
\begin{equation}
Y_{ij}^{u/d,\,\mathrm{UV}}  ~=~ 
\frac{3  \abs{g^{u/d}_{ij}}^2 A_*\, \mpl\, \ms}{64\pi^5}\, 
\int\limits_\xmin^\xmax \! \dd x\, (16 - 3x^2)\, K_2(x) + 8x\, K_1(x) +3 x^2\,
K_0(x)\;,
\label{eq:YsigmaUV}
\end{equation}
where $A_* \equiv 45/(2\pi^2 g_*^S  1.66 \sqrt{g_*^\rho})$. The
limits of integration are given by $\xmin = \ms/T_R$ and $\xmax =
\ms/T_\mathrm{EWPT}$, where $T_\mathrm{EWPT}$ is the temperature of the
EWPT, taken to be $T_\mathrm{EWPT} = 100~\gev$.  
Precise modeling of the dynamics of EWPT is beyond the scope of this work.

At small values of $\xmin$ the integrand in \Cref{eq:YsigmaUV} is
approximated by a power series, $\mathcal I(x) \approx (32/x^2) - 6 + \mathcal
O(x^2)$, while for $\ms \gtrsim 2\,T_\mathrm{EWPT}$, the integral is insensitive to
the precise value of $\xmax$. As a consequence it is reasonable to approximate
\begin{equation}
Y_{ij}^{u/d,\,\mathrm{UV}} ~\approx~ \frac{3  \abs{g^{u/d}_{ij}}^2 A_*\, \mpl \,T_R}{2\pi^5}
\label{eq:YsigmaUVapprox}
\end{equation}
in the range $T_\mathrm{EWPT} \ll \ms \ll T_R$. 

\subsection{Freeze{\tdash}In Below the Electroweak Phase Transition} \label{sec:yIR}

For a low reheating temperature, $T_R \lesssim \mathcal O(\tev)$, 
a significant contribution to the flavon yield comes from freeze{\tdash}in after the EWPT via the renormalizable interactions,
\Cref{eq:Lafter}. In particular, as long as
$\ms \geq (m_{i} + m_{j})$, where $m_{i}$ are the quark masses,
$2\rightarrow1$ scattering is the leading order flavon production mechanism. For
much smaller flavon masses $\ms < (m_{i} + m_{j})$ the $q\bar q
\rightarrow \sigma$ scattering is kinematically forbidden, and flavon production
is driven instead by $2\rightarrow 2$ scattering such as $q\bar q \rightarrow
\sigma g$. Here, we assume that the flavon is heavy enough that
the $2\rightarrow 1$ scattering is sufficient to describe low{\tdash}temperature flavon yield.

The infrared (IR) contribution to the flavon yield from a given $2\rightarrow 1$ scattering is
\begin{align}
Y^{(u/d),\,\mathrm{IR}}_{ij}~\approx~&\frac{3 A_*\, \mpl}{16\pi^3 \ms} 
\abs{ \tg_{ij}^{u/d}}^2  \left(1 - 2\frac{m_i m_j}{\ms^2} \right) \notag\\
&{}\times\sqrt{1-\frac{(m_i + m_j)^2}{\ms^2} } \sqrt{1-\frac{(m_i - m_j)^2}{\ms^2} }
 \int\limits_\xmin^\infty \! \dd x \, x^3 \,K_1(x)\;,
 \label{eq:YsigmaIR}
\end{align}
where $\xmin = m_\sigma/T_\mathrm{EWPT}$ and we effectively replace  
$\xmax \rightarrow \infty$ which is a good approximation provided $\ms\gtrsim\mathcal O(\gev)$. 
For low flavon masses, $\ms\lesssim T_\mathrm{EWPT}$, one can set
$\xmin\rightarrow 0$ and the remaining integral is equal to $3\pi/2$. 
The IR contribution to the yield is obtained after summing over all the relevant
processes $Y^\mathrm{IR}_\sigma = \sum_{ij}
Y_{ij}^{u,\,\mathrm{IR}}+\sum_{ij}Y_{ij}^{d,\,\mathrm{IR}}$. 
The contribution from the subprocess with top quarks in the initial state
is negligible since their number density is highly Boltzmann suppressed after the EWPT.

We also neglect the dynamical effects associated with the
EWPT by approximating a constant value for the Higgs VEV
and using the zero temperature flavon couplings for all $T<T_\mathrm{EWPT}=100~\gev$.

\subsection{Flavon Yield in the Early Universe}

\begin{figure}[t]
\centerline{\subfigure[~$Y_\sigma(T_R)$.\label{fig:yieldTR}]{%
\includegraphics[width=0.48\textwidth]{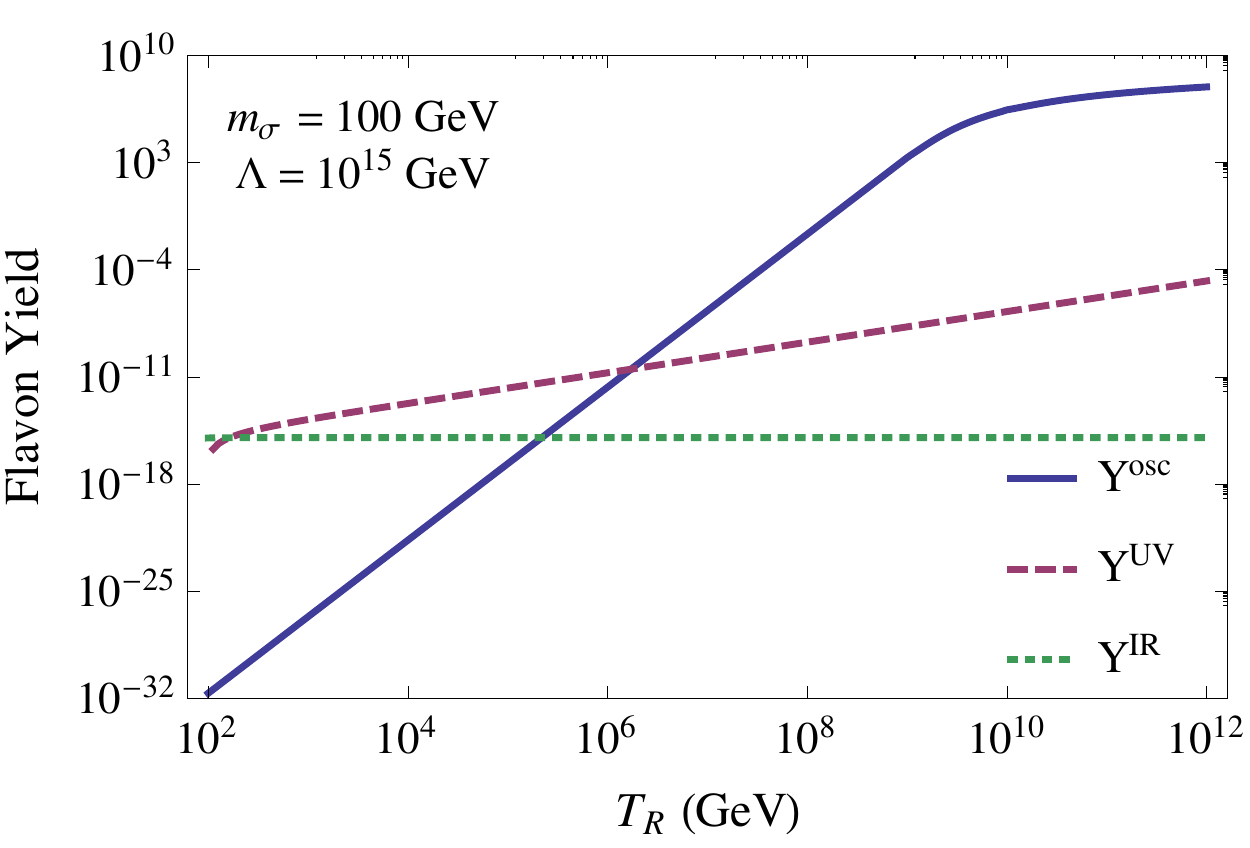}%
}\quad
\subfigure[~$Y_\sigma(m_\sigma)$.\label{fig:yieldmsigma}]{%
\includegraphics[width=0.48\textwidth]{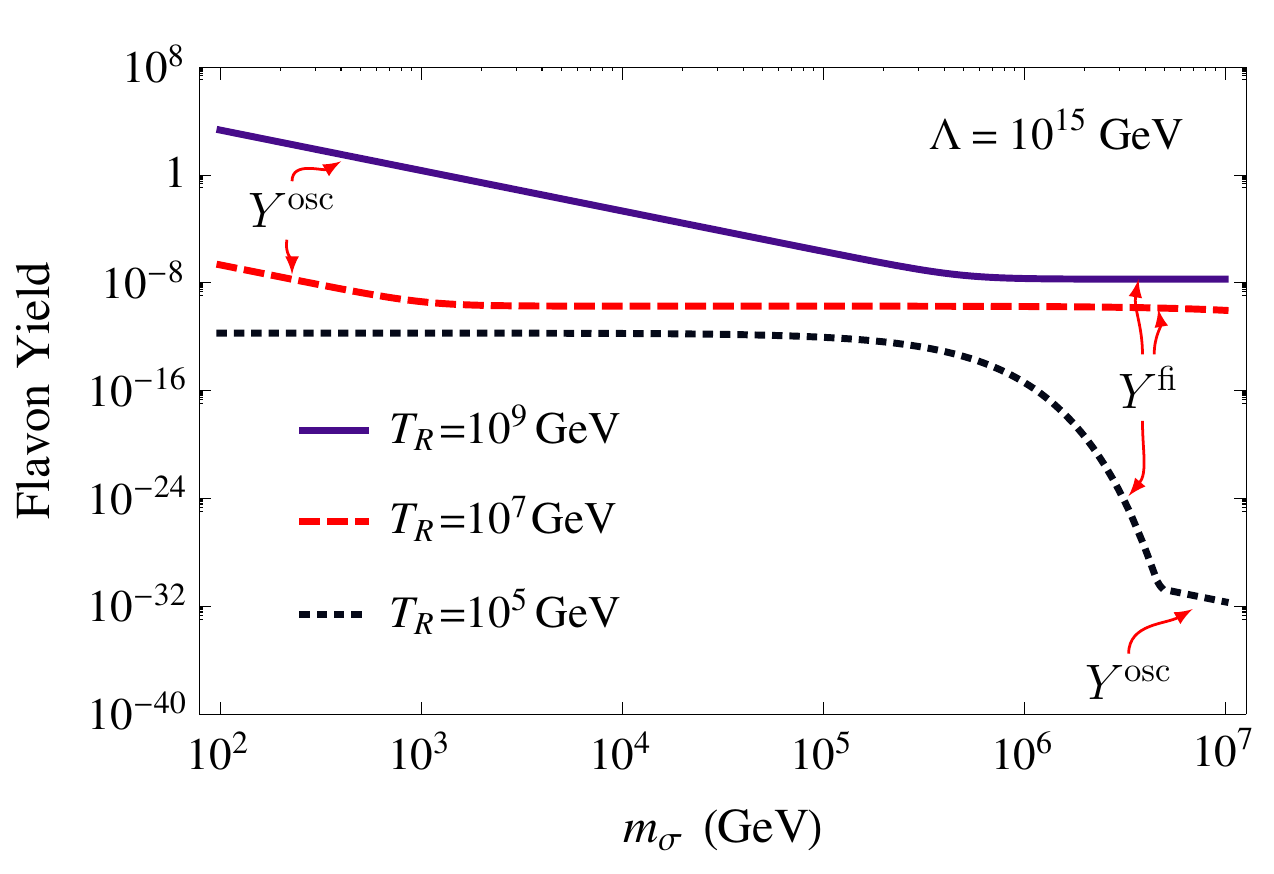}%
}}
\caption{The total flavon yield as (a) a function of $T_R$ for a fixed value of
$\ms=100\,\gev$ and (b) as a function of $\ms$ for fixed values of
$T_R=10^5, 10^7, 10^9\,\gev$. In both plots we take
$\Lambda=10^{15}~\gev$: all contributions to the flavon yield scale as
$\Lambda^{-2}$. In the right panel, we indicate whether the yield is driven
primarily by $Y^\mathrm{osc}$ or by the freeze{\tdash}in contribution $Y^\text{fi} =
Y^\text{UV} + Y^\text{IR}$. For $T_R = 10^5\; \gev$, the exponential suppression
of $Y^\text{fi}$ can be seen for $\ms > T_R$, while $Y^\mathrm{osc}$ obeys a
simple power law in this regime.%
}
\label{fig:yield}
\end{figure}

The total flavon yield is the sum of the contribution from flavon oscillations together with the IR and UV freeze{\tdash}in production:
\begin{equation}
Y_\sigma~=~\sum_{ij}{\left(Y_{ij}^{u,\,\mathrm{UV}}+Y_{ij}^{u,\,\mathrm{IR}}+
Y_{ij}^{d,\,\mathrm{UV}}+Y_{ij}^{d,\,\mathrm{IR}}\right)} + Y^\mathrm{osc}\;. \label{eq:finalyield}
\end{equation}
In \Cref{fig:yield}, we show the flavon yield $Y_\sigma$ as a function of $T_R$
and $\ms$, with flavon couplings that correspond to the parameter set
\eqref{eq:charges1} in \Cref{sec:flavoncouplings} for $\Lambda=10^{15}\,\gev$. All three contributions to
the yield scale uniformly as $\Lambda^{-2}$.

For large reheating temperatures, the freeze{\tdash}in yield is subdominant to the
oscillatory contribution $Y^\mathrm{osc}$, particularly for light flavons. The
situation is reversed at lower temperatures $T_R \ll \sqrt{\ms M_H}$, as can be
seen in \Cref{fig:yieldTR}. Here $Y^\text{UV}$ becomes the
largest contribution, provided $T_R$ and $\ms$ are still large compared to the
electroweak scale. This behavior is easily explained from
\Cref{eq:YsigmaUV,eq:Yosc}: the freeze{\tdash}in yield scales as $Y^\text{UV} \propto
T_R$, while the oscillatory component scales much more dramatically
as $Y^\mathrm{osc} \propto T_R^5$. 

For light flavons with $\ms \lesssim 100~\gev$, $Y^\text{IR}$ surpasses
$Y^\text{UV}$ for small reheating temperatures $T_R\lesssim \mathcal O(\tev)$,
but is otherwise negligible compared to $Y^\text{UV}$ and $Y^\mathrm{osc}$. 
This can be understood from \Crefrange{eq:YsigmaUV}{eq:YsigmaIR} in the limit of
$T_R\gg  T_\mathrm{EWPT}$ and large flavon mass,
\begin{equation}
Y^{(u/d),\,\mathrm{IR}}_{ij}~\lesssim~
\frac{v_h^2  }{\ms T_R} Y^{(u/d),\,\mathrm{UV}}_{ij} ~\ll~ 
Y^{(u/d),\mathrm{UV}}_{ij}\;.
\end{equation}

We close with a discussion of some of the effects not captured in our calculations:
\begin{itemize}
\item Our estimates treat the distribution functions according to 
Maxwell--Boltzmann statistics, which is questionable in the relativistic
$T\gg \ms$ limit.  To assess the error introduced by this simplification, we examine the impact of correcting the
statistical factors for one subprocess in \Cref{sec:statistics}, and find that it impacts the outcome by an $\mathcal O(1)$ factor.
\item The leading{\tdash}order freeze{\tdash}in yield is expected to be further modified when the effects from thermal quark masses 
$m_q^2(T) \sim \frac{1}{6} g_s^2 T^2$ are included (for a review, see for example~\cite{Laine:2016hma}). 
At high temperatures the effect is modest, but for light flavons at low temperatures $T_R \lesssim \tewk$ the 
thermal effects can restrict the phase space for the $2\rightarrow1$ scattering 
$q\bar q \rightarrow \sigma$ that determines $Y^\text{IR}$. We therefore restricted ourselves to scenarios with $T_R\gtrsim 1~\tev$.
\item There are corrections to the low-temperature freeze-in contribution from 
next{\tdash}to{\tdash}leading order QCD processes such as $q \bar q \rightarrow \sigma g$. 
These are enhanced by a color factor and are kinematically allowed even when $\ms < 2m_q(T)$. 
Even in the case of a light flavon with $\ms\lesssim 100~\gev$ and reheating temperature $T_R\sim 1~\tev$, the 
QCD corrections are expected to be not larger than an $\mathcal{O}(1)$ factor.
\end{itemize}
These simplifications and formally higher order corrections can be large, leading to an $\mathcal{O}(1)$
uncertainty in our results for the yield.  While beyond the scope of our work, it would be interesting to pursue refined
estimates for the flavon yield in the future.

\section{Cosmological Constraints on late{\tdash}time Decays of Flavons}
\label{sec:cosmoconstraints}

The flavons produced in the early universe will eventually decay back to SM
degrees of freedom.  Depending on the epoch in which this occurs, it
can have dramatic consequences for cosmology. For
$\ms\gtrsim 10\,\gev$, the primary concern is the possibility of spoiling
the successful predictions from big bang nucleosynthesis 
(for a similar analysis on light flavons see~\cite{Baldes:2016gaf}).
Late{\tdash}time energy injection from flavon decays could also cause
spectral distortions in the Cosmic Microwave Background (CMB)
radiation~\cite{Hu:1993gc,Chluba:2013wsa}, although these constraints are
typically less severe than the ones from BBN provided the decaying particle is heavy
enough. In addition, even if flavon decays happen earlier, they might cause a
significant entropy production that dilutes a primordial baryon asymmetry to
values inconsistent with observations. In the following, we discuss constraints on the flavon that
can be deduced from these processes.

\subsection{Flavon Decays}

The flavon lifetime depends on whether or not the electroweak symmetry is
broken. In particular, before the EWPT when the flavon interactions are described by
\Crefrange{eq:Lbefore}{eq:gbefore}, the decay width into $3${\tdash}body final state,
$\sigma\rightarrow \Phi\,\overline{Q}_i\,u_j$, is given by
\begin{eqnarray}
\Gamma_{ij}^{u/d,\,\mathrm{UV}} &=&
\frac{N_c}{3}\frac{|g_{ij}^{u/d}|^2}{64\, \pi^3}\frac{\ms^3}{\Lambda^2}\;.
\end{eqnarray}
On the other hand, if this decay rate is smaller than the Hubble rate at the
time of the EWPT, $\Gamma_{ij}^{u/d,\,\rm UV} < H(t_\mathrm{EWPT})$, the flavon
lifetime is effectively determined by the renormalizable couplings
\Crefrange{eq:Lafter}{eq:gafter}. The $2${\tdash}body flavon decay width into pairs of
quarks, $\sigma\rightarrow u_i\bar{u}_j+u_j\bar{u}_i$, then
reads~\cite{Bauer:2016rxs}
\begin{align}
\Gamma_{ij}^{u,\,\mathrm{IR}} ~=~& \frac{3\,\ms}{16\,\pi}\,\frac{v_\mathrm{EW}^2}{\Lambda^2}\,\left[\frac{\left[\ms^2-(m_i+m_j)^2\right]\left[\ms^2-(m_i-m_j)^2\right]}{\ms^4}\right]^{1/2}
\notag\\
 &{} \times\left\{\left(|\widetilde{g}^u_{ij}|^2+|\widetilde{g}^u_{ji}|^2\right)\left(1-\frac{m_i^2+m_j^2}{\ms^2}\right) - 4\,\Re(\widetilde{g}^u_{ij}\widetilde{g}^u_{ji})\,\frac{m_im_j}{\ms^2}\right\}
 \;,\label{GammaafterEWPT}
\end{align}
and similarly for down{\tdash}type quarks. In particular, as can be seen in
\Crefrange{eq:tildegupcharg1}{eq:tildegupcharg2}, flavon decays into a pair of
top quarks are possible when kinematically allowed. 
Note that a flavon coupling to a pair of top quarks 
is typically induced through the nontrivial basis change from $u_3$ and
$Q_3$ to the mass eigenstates $t_{\mathrm{L}/\mathrm{R}}$  (see
\Cref{eq:FlavonCouplingsEW1} and \Cref{eq:FlavonCouplingsEW2}).

The total flavon decay width $\Gamma_\sigma$ is the sum of the various flavor combinations
in the channels discussed above.
It determines the flavon decay
temperature, $T_\sigma$ from $\Gamma_\sigma\simeq H(T_\sigma)$,
\begin{equation}
T_\sigma~=~\left(\frac{90}{\pi^2\,g_{\ast,\rho}}\right)^{1/4}\,
\sqrt{\mpl\,\Gamma_\sigma}
~
\widesim{\mathrm{IR}}~0.2\cdot \frac{v_\mathrm{EW}}{\Lambda}\,\sqrt{\mpl\ms}\;.
\label{eq:Tsigma}
\end{equation}
An additional suppression of $T_\sigma$ on top of the simple scaling in
\Cref{eq:Tsigma} is obtained for $\ms<m_t$ since then the dominant decay
channels of flavon involving the top quark are forbidden.
For cases in which the flavon temporarily
dominates the energy density of the universe, $T_\sigma$ is the temperature at
which the radiation dominated epoch is restored when the flavon decays.

\subsection{Big Bang Nucleosynthesis Constraints}
\label{sec:BBN}

If the lifetime of the flavon, $\tau_\sigma$, exceeds $\sim 0.1-1\,\sec$,
its decays are subject to potentially stringent constraints from its potential
to destroy the successful predictions of the BBN (for recent reviews
see Refs~\cite{Iocco:2008va,Cyburt:2015mya}), e.g. by producing
electromagnetic and/or 
hadronic showers which destroy light nuclei (see,
e.g.,~\cite{Kawasaki:1994af,Cyburt:2002uv,Kawasaki:2004yh,Kawasaki:2004qu,Jedamzik:2006xz}
and references therein).
As a result, there are bounds from BBN on the yield of sufficiently
long{\tdash}lived flavons:
\begin{itemize}
\item For large enough $Y_\sigma$ and $\tau_\sigma\lesssim 10^2\sec$, hadronic
particles produced in $\sigma$ decays can cause $p\leftrightarrow n$
interconversion, affecting the $^4\text{He}$ mass fraction, $Y_p$, by
changing the neutron{\tdash}to{\tdash}proton ratio. 
At later times, for $\tau_\sigma\gtrsim
10^2\sec$, injected high{\tdash}energy hadrons are no longer effectively stopped in
thermal plasma~\cite{Kawasaki:2004qu} and hadrodissociation processes take
place. As a result, the BBN constraints become much stronger in this regime and
typically come from deuterium overproduction caused by hadrodissociation of
$^4\mathrm{He}$. 
\item Electromagnetic cascades initiated by decays of flavons typically get
equilibrated by scatterings of injected high{\tdash}energy $\gamma${\tdash}rays on
background photons that lead to $e^+e^-$ pair production. However, 
for lower temperatures, pair production becomes inefficient and
photodissociation of $\mathrm{D}$ or even $^4\mathrm{He}$ can occur for
$\tau\gtrsim 10^4\sec$ and $\tau\gtrsim 10^6\sec$, respectively
leading to constraints comparable to those from hadronic decays.
\end{itemize}
Constraints on late{\tdash}decaying particles from BBN have recently been
revisited~\cite{Kawasaki:2017bqm} to include up{\tdash}to{\tdash}date
observational data concerning the primordial abundances of light elements
and improved determinations of the rates of relevant nuclear reactions. 
Particularly relevant are new stringent limits on the $\text{D}/\text{H}$ ratio
based on significantly reduced observational uncertainties~\cite{Cooke:2013cba}
and updates to $Y_p$ incorporating the recent study of emission lines in both infrared and
visible wavelengths in $45$ extragalactic HII 
regions\footnote{This measurement was initially claimed to be inconsistent
with standard BBN predictions~\cite{Izotov:2014fga}, but more recent analysis
indicates no statistically meaningful discrepancy~\cite{Aver:2015iza}.}~\cite{Izotov:2014fga}.

\begin{figure}[t]
\centerline{
\subfigure[~High $T_R$.\label{fig:HighTR}]{
\includegraphics[width=0.48\textwidth]{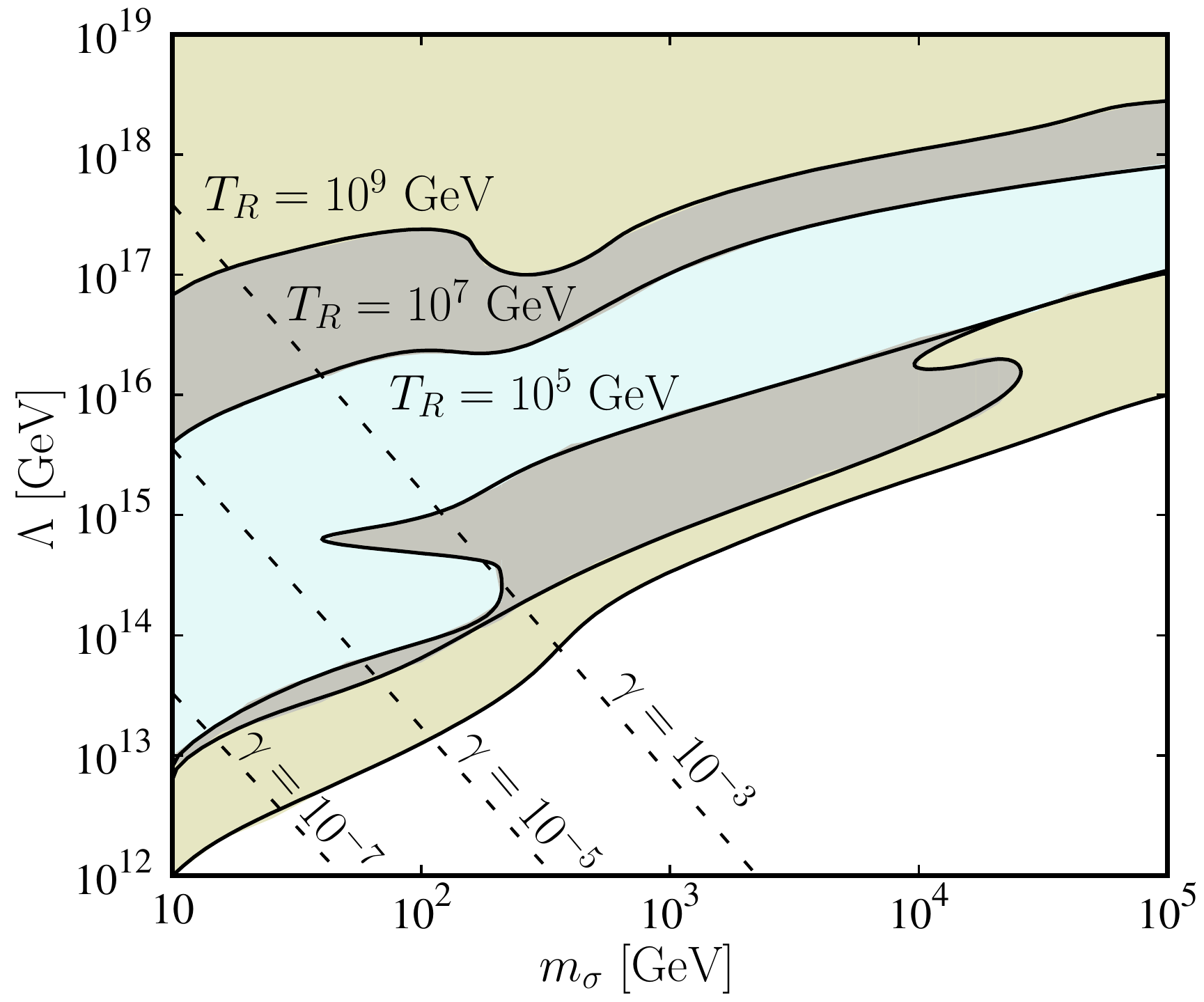}
}
\quad
\subfigure[~Low $T_R$.\label{fig:LowTR}]{
\includegraphics[width=0.48\textwidth]{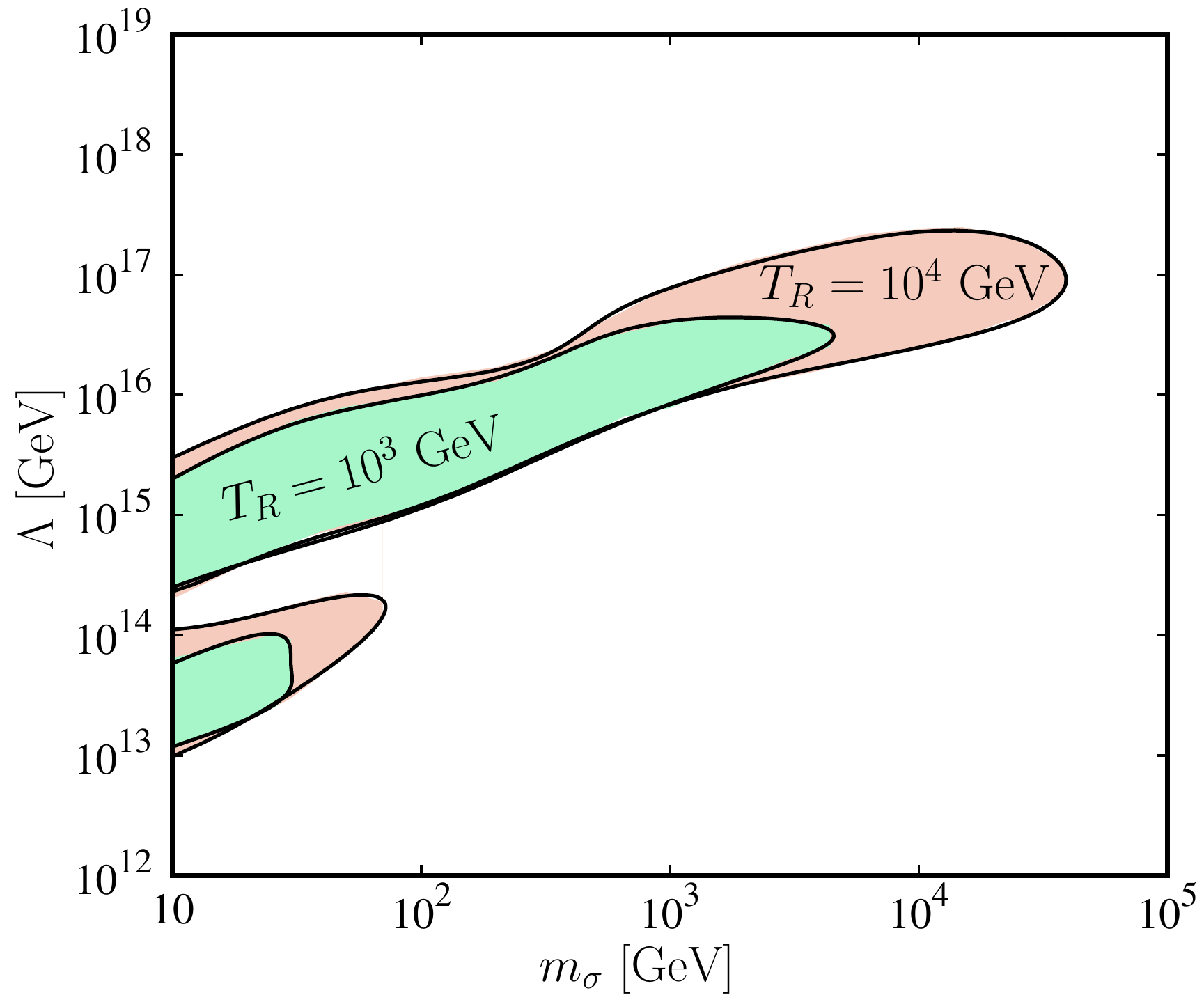}
}
}
\caption{Big Bang Nucleosynthesis constraints on late{\tdash}time flavon decays in the
$(\ms,\Lambda)$ plane. In (a) the light blue (gray, olive) shaded
areas correspond to regions in the parameter space excluded for $T_R=10^5\,\gev$
($10^7\,\gev$, $10^9\,\gev$). In (b) the limits are shown for
$T_R=10^3\,\gev$ and $10^4\,\gev$ as green and red shaded areas, respectively.}
\label{fig:BBNexcl}
\end{figure}

In the left panel of 
\Cref{fig:BBNexcl} we show the BBN constraints on the $(m_\sigma,\Lambda)$
plane for several values of the reheating temperature from $T_R = 10^5$ to $10^9$~GeV. 
For a given $\ms$, the BBN limits exclude a window of
$\Lambda$ starting where the flavon lifetime corresponds to the onset of
BBN, $\tau_\sigma\gtrsim 0.1-1\,\sec$ and
ending when $\Lambda$ becomes large enough that the flavon yield
decreases to the point that its decays become irrelevant for BBN.
For $T_R=10^9\,\gev$, this upper boundary
occurs for $\Lambda>\mplanck$. 
The detailed shape of the exclusion regions for $m_\sigma \leq 100$~GeV correspond to the interplay of
multiple observables, with the lower
$\Lambda$ (\ie, lower $\tau_\sigma$) constraints dominated by
 $p\leftrightarrow n$ interconversion and the larger values excluded by
hadrodissociation processes.
For $T_R=10^3\,\gev$ to $10^4\,\gev$ the situation is qualitatively the
same, as can be seen in the right panel of \Cref{fig:HighTR}. A new feature
for these lower reheat temperatures
is an intermediate range of $\Lambda\sim 10^{14}\,\gev$ for
which the flavon yield is already low enough to avoid the limits from
$p\leftrightarrow n$ interconversion processes, while its lifetime is short
enough that the constraints from hadrodissociation do not apply, resulting in a channel of 
allowed parameter space.

If the reheating temperature exceeds the freeze{\tdash}out temperature,
\Cref{eq:Tfo}, the flavon could enter thermal equilibrium at some stage of the
cosmological evolution and might dominate the energy density of the universe at later times
when it becomes nonrelativistic. In this case, we require
$T_\sigma>T_\mathrm{BBN}$ in order to prevent the early stages of BBN from proceeding during
a matter dominated epoch.

The exclusion bounds shown in \Cref{fig:BBNexcl} 
depend on the choice of FN charges $n_{ij}^{u/d}$ and the associated
order one coefficients $y_{ij}^{u/d}$. 
For natural models realizing the FN mechanism,
the variations in the flavon couplings are of order one (see \Cref{sec:flavoncouplings}). 
Thus, the model-dependence of the bounds is typically at most of the same order as the
theoretical uncertainties discussed in \Cref{sec:flavonproduction}. 
We verify that the two example
choices of FN charges \eqref{eq:charges1} and \eqref{eq:charges2}
result in exclusion bounds which differ by a factor of order unity, and choose to simplify
the presentation by focusing on the assignment \eqref{eq:charges1}.

As noted in~\cite{Kawasaki:2017bqm}, if one adopts the initial determination of
the $^4\mathrm{He}$ mass fraction from~\cite{Izotov:2014fga}, the resulting
discrepancy with respect to the standard BBN predictions can be explained by an
additional hadronic energy injection from decays of a new heavy particle with
a lifetime of $\tau\lesssim 10^2~\sec$ (in order to avoid exclusion bounds from
hadrodissociation processes), whose yield is close to the upper limit from
$p\leftrightarrow n$ interconversions. Interestingly, this scenario can be
realized by late{\tdash}time decays of the weakly coupled flavon. For
$\ms=1~\tev$ (used for illustration purposes in~\cite{Kawasaki:2017bqm}),
the discrepancy disappears provided 
$\Lambda\sim 10^{15}\,\gev$ and $T_R\sim (10^6-10^7)~\gev$.

\subsection{Dilution of the Baryon Asymmetry of the Universe}
\label{sec:baryondilution}

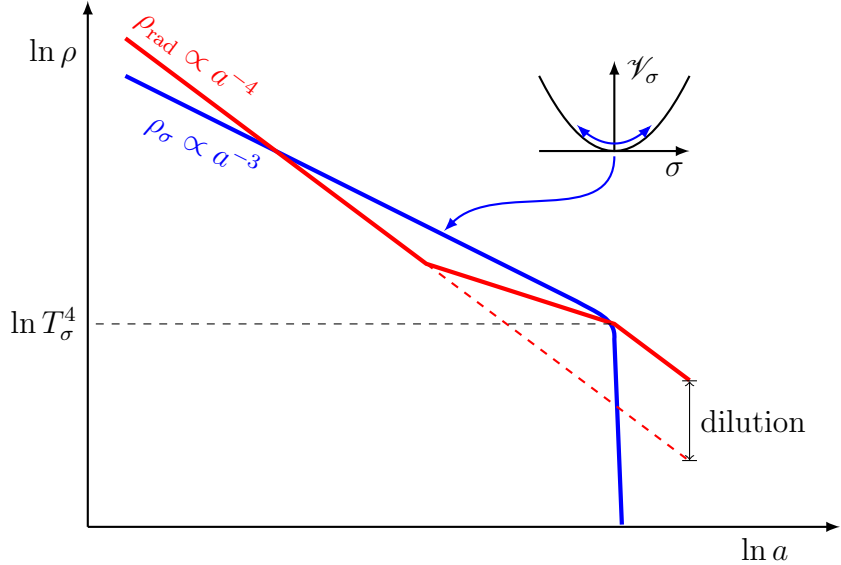
\begin{figure}[t]
\begin{center}
\begin{tikzpicture}
\draw[thick,-latex] (0,0) -- (10,0) node[pos=0.9,below]{$\ln a$};
\draw[thick,-latex] (0,0) -- (0,7) node[pos=0.9,left]{$\ln \rho$};
\draw[ultra thick,blue] (0.5,6) -- (6.5,3) 
node[pos=0.2,sloped,below]{$\rho_\sigma\propto a^{-3}$} 
coordinate[pos=0.7] (x)
to[out=-30,in=85] (7,2.5) -- (7.1,0.03);
\draw[ultra thick,red] (0.5,6.5) -- (4.5,3.5) 
node[pos=0.18,sloped,above]{$\rho_\mathrm{rad}\propto a^{-4}$} 
-- (7,2.7) -- (8,1.95) coordinate (a);
\draw[thick,red,dashed] (4.5,3.5) -- (8,{0.5+0.5*0.75}) coordinate (b);
\draw[|<->|] (a) -- (b) node[midway,right]{dilution};
\draw[dashed] (6.96,2.7) -- (0,2.7) node[left]{$\ln T_\sigma^4$};
\begin{scope}[xshift=7cm,yshift=5cm]
\draw[thick,-latex] (-1,0) -- (1,0) node[pos=0.9,below]{$\sigma$};
\draw[thick,-latex] (0,0) coordinate(O) -- (0,1.2) node[pos=0.9,right]{$\mathscr{V}_\sigma$};
\draw[thick] plot[variable=\x,domain=-1:1] ({\x},{\x*\x});
\draw[thick,blue,latex-latex] plot[variable=\x,domain=-0.5:0.5] ({\x},{\x*\x+0.1});
\end{scope}
\draw[thick,blue,-latex] ([yshift=-2pt]O) to[out=-90,in=45] ([xshift=1pt,yshift=1pt]x);
\end{tikzpicture}
\end{center}
\caption{Cartoon of the moduli problems associated with weakly coupled flavons.}
\label{fig:FlavonModuliProblem}
\end{figure}

In cases where the flavon dominates the energy density of the universe, its subsequent
decays can produce significant entropy and effectively wash out
baryons, leading to a reduced value of the primordial baryon asymmetry,
as illustrated schematically in \Cref{fig:FlavonModuliProblem}.
In this regime, the flavon yield is typically dominated by the contribution from 
field oscillations described in \Cref{eq:Yosc}.  We assume that the flavon does not reach
equilibrium, $T_\mathrm{dec} > T_R$.
The respective dilution factor (see \Cref{fig:FlavonModuliProblem} for an
illustration), $\gamma \leq 1$, is given by the ratio between
the radiation entropy density at $T_\sigma$ calculated as if there were no
flavon, $s_\mathrm{before}$, and the radiation entropy density after the flavon
decay, $s_\mathrm{after}$ (see, \eg,~\cite{Ema:2018abj})
\begin{equation}
 \gamma~ \equiv ~\frac{s_\mathrm{before}}{s_\mathrm{after}}
 ~\simeq~\min\left[\frac{3}{4}T_\sigma\,
 \frac{s^\mathrm{ini}}{\rho_\sigma^\mathrm{ini}},1\right]
 ~=~
 \min\left[\frac{3\,T_\sigma}{4\,m_\sigma\, Y_\sigma^\mathrm{ini}},1\right]\;,
\label{eq:dilution}
\end{equation}
where $\rho_\sigma^\mathrm{ini}$, $s^\mathrm{ini}$ and $Y_\sigma^\mathrm{ini}$
are the flavon energy density, entropy density and flavon yield evaluated at a
time when the initial amplitude of $\sigma$ field oscillations was set (see
\Cref{sec:oscillations} for details).
Typical values of $\gamma$ are shown  in \Cref{fig:HighTR} in the $(\ms,\Lambda)$
plane for $T_R=10^9~\gev$.

The precise impact of flavon decays on the baryon asymmetry depends on the specific model
of baryogenesis, in particular on the temperature at which the baryon asymmetry
is generated.
Scenarios such as thermal leptogenesis~\cite{Fukugita:1986hr}, for which the baryon
asymmetry is at best linear in $T_R$, are difficult to reconcile with weakly
coupled flavons, since $Y^\mathrm{osc}_\sigma\propto T_R^5$ (cf.~\Cref{eq:Yosc}).
In particular, for the maximal CP violation, the baryon-to-entropy ratio in the
thermal leptogenesis scenario is given by~\cite{Ema:2016ops}
\begin{equation}
\frac{n_b}{s} \lesssim \gamma \times 6\times 10^{-11}\,\times\left(\frac{m_{N_1}}{10^{11}~\gev}\right)\;,
\label{eq:thermlepto}
\end{equation}
where $m_{N_1}$ is the mass of the right-handed neutrino. As can be seen in
\Cref{fig:HighTR}, a light flavon with
$\Lambda\sim 10^{12}~\gev$ (at which scale it does not equilibrate for
$T_R=10^9~\gev$) is characterized by very small values of $\gamma$. In this
case, the observed value of $n_b/s\simeq 9\times 10^{-11}$~\cite{Ade:2015xua}
can only be reproduced for unreasonably large values of $m_{N_1}$.
This constraint is complementary to the bounds from BBN discussed in \Cref{sec:BBN}.

\subsection{Implications for Model Building}

While our results are derived in the specific framework of FN models, our
discussion of thermal corrections to the flavon potential applies to a more
general class of models where the quark Yukawa couplings
are field-dependent. Such scenarios include string and supergravity
models, in which the Yukawa couplings are moduli{\tdash}dependent, and $\sigma$ can
stand for any of the moduli, as well as bottom{\tdash}up models in which flavor is encoded through dynamics.

As derived in \Cref{sec:FreeEnergyYukawa}, the free energy becomes
smaller for decreasing Yukawa couplings, and thermal corrections tend to drive the
effective Yukawa couplings smaller at high temperatures. This is analogous to the statement that thermal
corrections tend to switch off the gauge
couplings~\cite{Buchmuller:2003is,Buchmuller:2004xr}.

What are the implications of these effects? As far as string model building is
concerned, we find that the so{\tdash}called moduli
problem~\cite{Coughlan:1983ci,deCarlos:1993wie} is often worse than appreciated. 
It has been argued (see e.g.~\cite{Kofman:2004yc} and references
therein) that the oscillations of moduli are suppressed if the moduli become
trapped at symmetry{\tdash}enhanced points. However, we find that if the
gauge and/or Yukawa couplings depend on these moduli, they get pushed away from
their $T=0$ minimum by an amount given in \Cref{eq:DeltaSigma}, even if
trapped there during inflation. We stress that these amplitudes are
model{\tdash}independent.

Let us close with a comment on other scenarios with varying Yukawa couplings. It
has been argued (see e.g.~\cite{Berkooz:2004kx}) that in the SM (extended by an
FN sector) there might be a first order phase transition if at temperatures of
the order $T_\mathrm{EW}$ the flavon has an expectation value of the order
$\Lambda$. However, we find that thermal corrections drive the flavon to smaller
values. It would thus be interesting to assess the viability of such scenarios with
the thermal corrections taken into account.

\section{Conclusions}
\label{sec:conclusions}

We have studied the phenomenology of models in which the SM Yukawa
couplings are set by the VEV(s) of some field(s), $\sigma$. Such fields are
referred to as flavons or moduli in the literature. When these flavons are
weakly coupled, they evade detection by collider experiments, but can nevertheless produce
large effects in cosmology.
Along the way, we have uncovered an important disagreement with previous analyses stating that the
flavon coupling to a pair of top quarks vanishes. Rather, we find that it is often rather sizable as
detailed in \Cref{sec:flavoncouplings}.

In the early universe, flavons are produced either in scatterings of the SM particles,
or result from coherent
oscillations triggered by thermal deformations of their potential. While the first
production mode is very similar to the one of gravitinos, axinos and other
ultraweakly coupled particles, to our knowledge the second production mode has not been
discussed in the literature and we find that it dominates in large portions of the
parameter space of interest.

Our analysis provides some key lessons. First, even if the couplings of the
flavon are suppressed by a mass scale $\Lambda$ that is large enough to render
collider experiments irrelevant, cosmology nonetheless places restrictions on
the mass of the flavon. If $\ms$ is rather small, late decays of the flavon may
spoil BBN. Even if the flavon decays before BBN, it may  (depending on the
specific model of baryogenesis) unacceptably dilute the baryon asymmetry of the
universe. Both of these issues have been previously discussed in the context of
the so{\tdash}called moduli problem. However, our results cast doubt on some of the
proposed solutions to this problem when the Yukawa couplings of the SM fermions
depend on the  modulus' in question. In more detail, it is not sufficient to
provide a mechanism that fixes the moduli at the minimum of the
zero{\tdash}temperature potential during inflation, since they will still typically
get pushed away after reheating and thus jeopardize cosmology.

It is also worth mentioning that the finite temperature corrections to the potential are
such that flavons get driven to values at which the Yukawa couplings are
\emph{smaller}. It will thus be interesting how scenarios in which the Yukawa
couplings are taken to be larger in the early universe get modified after our
model{\tdash}independent corrections have been taken into account. 

Beyond the specifics
of theories in which a modulus controls the SM notion of flavor, 
our results illustrate the importance of cosmology as a window into particle physics
at energy scales far beyond the reach of current collider or conventional astrophysical probes.

\acknowledgments

We are grateful to Mu{\tdash}Chun Chen, Tilman Plehn, and Arvind Rajaraman for
insightful discussions, and Marco Drewes for correspondence. This work is
supported in part by NSF Grant No.~PHY-1620638 and PHY-1719438. S.T.\ is
supported in part by the Polish Ministry of Science and Higher Education under
research grant 1309/MOB/IV/2015/0 and by the National Science Council (NCN)
research grant No.\ 2015-18-A-ST2-00748.  

\newpage
\appendix

\section{Realistic Flavon Couplings}
\label{sec:flavoncouplings}

We consider two representative choices of the FN charges that are commonly used in the
literature~\cite{Binetruy:1996xk,Berkooz:2004kx}:
\begin{subequations}
\begin{align}
Q^{(1)}_\mathrm{FN}(\overline{Q}_{i})& = (3,2,0)\;,~ Q^{(1)}_\mathrm{FN}(u_{i})
= (5,2,0)\;,~ Q^{(1)}_\mathrm{FN}(d_{i}) = (4,3,3)\;,
\tag{FN charges 1}\label{eq:charges1}\\
Q^{(2)}_\mathrm{FN}(\overline{Q}_{i}) &= (3,2,0)\;,~ 
Q^{(2)}_\mathrm{FN}(u_{i}) = (4,1,0)\;,~ Q^{(2)}_\mathrm{FN}(d_{i}) = (3,2,2)\;.
\tag{FN charges 2}\label{eq:charges2}
\end{align}
\end{subequations}
Both sets differ by $\UFN$ charges assigned to $\text{SU}(2)_\mathrm{L}$ singlet
fields with the first choice corresponding to systematically larger values, but
are consistent with observations for an appropriate 
choice of the $\mathcal{O}(1)$ coefficients $y^{u/d}_{ij}$ in \Cref{eq:L}.

In both
scenarios the FN charges of the up{\tdash}type third generation quarks vanish,
$Q_\mathrm{FN}(Q_3)=Q_\mathrm{FN}(u_3)=0$, as dictated by the top
Yukawa coupling, $(Y_u)_{33}\sim 1$. This, in turns, implies that before EWPT
there are no tree{\tdash}level flavon interactions with a pair of top quarks. On the
other hand, as we shall see in more detail below, after EWPT a nonzero $\sigma
t\bar t$ coupling gets induced upon rotation to the mass basis. Although this
effect can be non-negligible, the dominant flavon couplings are typically the
off{\tdash}diagonal $\sigma t\bar c$ and the diagonal $\sigma b \bar b$. In particular, in the
flavor basis they correspond to $\sigma\, \overline{Q}_3\,u_2$ and
$\sigma\,\overline{Q}_3\,d_3$ couplings, respectively, \ie, they are
proportional to $\varepsilon$ ($\varepsilon^2$) and $\varepsilon^2$
($\varepsilon^3$) for the first (second) set of the FN charges. 

While these two charge assignments lead to flavon couplings
with parametrically different dependence on $\varepsilon$, this difference
is partially compensated by the $y_{ij}^{u/d}$ necessary to reproduce
observations.
We determine
these coefficients by performing a fit to the values~\cite{Patrignani:2016xqp}
of the quarks masses, the CKM matrix entries and the Jarlskog invariant with the
use of the Multinest scanning tool~\cite{Feroz:2008xx}.  The Yukawa couplings
are evolved under the renormalization group using the
SPheno~\cite{Porod:2003um,Porod:2011nf} implementation of the SM
via the SARAH package~\cite{Staub:2013tta}. Below we
present the results for $y^{u/d}_{ij}$ at high{\tdash}energy scale $10^{10}~\gev$. 
With those coefficients determined, the flavon couplings above EWPT can
then be obtained from \Cref{eq:gbefore}. 
The couplings at and below the scale of EWPT, $\widetilde{g}^{u/d}$, are obtained upon rotation to the
mass basis as shown in \Cref{eq:gafter}.

Since our aim is to explore realistic model points and examine how these two
commonly used charge assignments differ in terms of their cosmological bounds,
we keep only the leading digits in the fit points and leave careful exploration of the allowed ranges
of parameters for future work.
In our numerical studies, it turns out that the differences in the cosmological
constraints from both sets of charges and flavon couplings are 
never larger than a factor of a few, and thus have minor impact on the
exclusion plots.  Thus, in \Cref{sec:flavonproduction,sec:cosmoconstraints}
we present results based on \eqref{eq:charges1}.

Our fit determines:
\begin{itemize}
\item  For the choice~\eqref{eq:charges1}, the $\mathcal{O}(1)$
coefficients at the scale $10^{10}\,\gev$ are
\begin{subequations}
\begin{align}
y^u ~\simeq&~ \begin{pmatrix}
0.3 + 0.3\,\I & -0.2 + 0.5\,\I & 0.2 -0.2\,\I\\
-0.3+0.2\,\I & -0.4+0.5\,\I & 0.5-0.3\,\I\\
0.4-0.3\,\I & -0.4-0.9\,\I & 0.3+0.5\,\I
\end{pmatrix}\;,\\
y^d ~\simeq&~ \begin{pmatrix}
0.4 +0.2\,\I & 0.2 -0.2\,\I & 0.2 -0.3\,\I\\
0.1 -0.6\,\I & 1.0-0.1\,\I & -0.2-0.4\,\I\\
-0.4+0.3\,\I & -0.4+0.6\,\I & 0.4 + 0.5\,\I
\end{pmatrix}\;.
\end{align}
\end{subequations}
The flavon couplings at the scale of EWSB in the quark mass basis are
\begin{subequations}\label{eq:FlavonCouplingsEW1}
\begin{align}
\widetilde{g}^u~\simeq&~\frac{v_\mathrm{EW}}{2\,\Lambda}\,\begin{pmatrix}
(-4 - 0.2\,\I)\times 10^{-4} & (0.9- \I)\times 10^{-3} & (7-2\,\I)\times 10^{-2}\\
(-0.6- 5\,\I)\times 10^{-4} & (0.1- 50\,\I)\times 10^{-3} & 0.3+0.5\,\I\\
(2-3\,\I)\times 10^{-3} & 0.7-0.4\,\I & 0.09 -0.05\,\I
\end{pmatrix}\;,
\label{eq:tildegupcharg1}\\
\widetilde{g}^d~ \simeq&~\frac{v_\mathrm{EW}}{2\,\Lambda}\,\begin{pmatrix}
(-2-4\,\I)\times 10^{-4} & (0.5 -3\,\I)\times 10^{-4} & (-2+2\,\I)\times 10^{-3}\\
(-1 -6\,\I)\times 10^{-4} & (-10+ 2\,\I)\times 10^{-3} & (-7-7\,\I)\times 10^{-3}\\
(7-7\,\I)\times 10^{-3} & (-2-2\,\I)\times 10^{-3} & -0.2 + 0.2\,\I
\end{pmatrix}\;.
\label{eq:tildegdowncharg1}
\end{align}
\end{subequations}
\item For the choice~\eqref{eq:charges2}, the $\mathcal{O}(1)$ coefficients at
 the scale $10^{10}\,\gev$ are
\begin{subequations}
\begin{align}
y^u~\simeq&~ \begin{pmatrix}
0.4 + 0.4\,\I & 0.1 + 0.1\,\I & 0.2 - 0.2\,\I\\
-0.1 + 0.3\,\I & 0.1 + 0.2\,\I & 0.4 - 0.5\,\I\\
0.8 - 0.1\,\I & -0.3 - 0.5\,\I & 0.2 + 0.5\,\I
\end{pmatrix}\;,
\\
y^d~\simeq&~ \begin{pmatrix}
0.5 + 0.1\,\I & 0.2 - 0.1\,\I & 0.2 - 0.1\,\I\\
0.6 - 0.1\,\I & 0.5 - 0.1\,\I & -0.1 - 0.2\,\I\\
-0.1 + 0.1\,\I & -0.1 + 0.1\,\I & 0.1 + 0.1\,\I
\end{pmatrix}\;.
\end{align}
\end{subequations}
The flavon couplings at the scale of EWSB in the quark mass basis read
\begin{subequations}\label{eq:FlavonCouplingsEW2}
\begin{align}
\widetilde{g}^u ~\simeq&~ \frac{v_\mathrm{EW}}{2\,\Lambda}\,\begin{pmatrix}
( -1+9\,\I)\times 10^{-4} & (4+5\,\I)\times 10^{-3} & (-40 + 2\,\I)\times 10^{-2}\\
(-40-4\,\I)\times 10^{-4} & 0.02 +0.2\,\I & -0.5 +0.2\,\I\\
(50-6\,\I)\times 10^{-3} & -0.5 -\I & -0.3+ 0.2\,\I
\end{pmatrix}\;,\label{eq:tildegupcharg2}\\
\widetilde{g}^d ~\simeq&~ \frac{v_\mathrm{EW}}{2\,\Lambda}\,\begin{pmatrix}
(-30 -3\,\I)\times 10^{-4} & (10-7\,\I)\times 10^{-4} & (5 -6\,\I)\times 10^{-3}\\
(5+7\,\I)\times 10^{-4} & 0.01 -0.02\,\I & -0.02+0.2\,\I\\
(20 -8\,\I)\times 10^{-3} & (-3 -4\,\I)\times 10^{-3} & 0.2 +0.2\,\I
\end{pmatrix}\;.
\label{eq:tildegdowncharg2}
\end{align}
\end{subequations}
\end{itemize}

We note that in contrast to statements in the literature, the 3--3 entries in
the couplings of $\widetilde{g}^u$, i.e.\ the couplings of the flavon to the $t$
quark does not vanish. Rather, using the values of \cite{Bauer:2016rxs} we
obtain a coupling of magnitude $0.1$ (see \Cref{eq:tildegupcharg1}) while for
the choice \eqref{eq:charges2} we obtain couplings larger than $0.3$ (see
\Cref{eq:tildegupcharg2}).  These results are not optimized, and it may be
possible to arrive at larger or smaller couplings to the top quark for different
sets of the $\mathcal{O}(1)$ coefficients.

\section{\boldmath Lifting the $\rho$ Mass\unboldmath}
\label{sec:scalarpotential}

In this appendix, we present a model which lifts the pseudoscalar $\rho$ from the spectrum
without additional distortion to the phenomenology.  If the global symmetry
$\UFN$ were to be exact, $\rho$ would be a Goldstone boson, and thus massless.  
On the other hand, if one were to
explicitly break the $\UFN$ symmetry such that $\rho$ acquires a mass,
this would generically allow for operators which contribute to the SM Yukawa interactions and
destroy the FN mechanism.

One option could be to
gauge $\UFN$, so that $\rho$ is eaten as the longitudinal mode of the associated
gauge boson, but in that case the massive gauge boson itself complicates
the discussion. 
For an unsuppressed gauge coupling, the gauge
boson would be heavy, but would mediate additional interactions between the flavon and the 
SM matter fields. For a very small gauge coupling, these interactions would be
suppressed, but the gauge boson would be light, and could play an important role in low
energy phenomenon.

\medskip

A better option is to replace $\UFN$ by a discrete
$\mathbbm{Z}_N$ symmetry, where $N$ is even and larger than twice the largest
Froggatt--Nielsen charge. The charge assignment for the $\mathbbm{Z}_N$
coincides with the one of $\UFN$. In this case, the flavon potential
\eqref{eq:VS} is given by
\begin{align}
 \mathscr{V}_S&~=~-\mu_S^2\,S^*S+\frac{\lambda}{4}\left(S^*S\right)^2
 +\frac{\kappa}{\Lambda^{N-4}}\left(S^N+(S^*)^N\right)
 +\frac{\eta}{\Lambda^{N-2}}\left(S^*S\right)^{1+N/2}\;,
\end{align}
where we omit other higher order couplings and $\kappa$ and $\eta$ are
dimensionless parameters. The presence of the $\kappa$ term shifts the VEV of the flavon, 
and more importantly leads to a mass term for the
would{\tdash}be{\tdash}Goldstone mode $\rho=\im S/\sqrt{2}\,$,
\begin{align}
 m_\rho^2&~\sim~\frac{v_S^N}{\Lambda^{N-2}}\;,
 \quad\text{where}~v_S~\sim~\sqrt{\frac{\mu_S^2}{\lambda}}\;.
\end{align}
As discussed in the main text, we consider small $\lambda$. At the
same time, we want the flavon to settle around $0.23\,\Lambda$, so
$\mu_S^2\ll\Lambda^2$. In this regime the $\UFN$ breaking terms can be more important
than the  $\UFN$ conserving terms. For example, for the parameters 
\[ \Lambda~=~10^{10}\,\gev\;,\quad
N~=~16\;,\quad \lambda~=~10^{-9}\;, \quad
\kappa~=~  0.008 \quad\text{and} \quad\eta~=~0.4\;,
\]
we obtain a minimum at $\re S=0.23\cdot\Lambda$  with
\[ m_{\sigma}^2~\simeq1.67\cdot 10^{11}\,\gev^2
\quad \text{and}\quad
m_\rho^2~\simeq~2.37\cdot 10^{11}\,\gev^2\;,\]
provided that $\mu_S^2=1.78\cdot10^{10}\,\gev^2$,
which realizes the desired pattern of masses.

\section{Free Energy of a Relativistic Gas of Fermions and Bosons with Yukawa Interactions}
\label{sec:FreeEnergyYukawa}

In this appendix, we compute the thermal correction to the free energy due to
Yukawa couplings.  To this end, we consider a massless Dirac fermion $\Psi$, a
massless complex scalar $\phi$, and the interaction Lagrangian
\begin{align}
 \mathscr{L}_\mathrm{Yukawa}&~=~-y\,\overline{\Psi}\,P_\mathrm{L}\,\Psi\,\phi
 +\text{h.c.}\;.
\end{align}
The free energy of a relativistic gas of fermions and bosons with Yukawa
interactions can be computed using basic methods of thermal field theory. 
Following the discussion of \cite{Laine:2016hma} around Equation~(5.116), we
obtain the leading{\tdash}order correction to the free energy as
\begin{align}\label{eq:DeltaF1}
 \Delta \mathcal{F}&~=~\CenterObject{\includegraphics{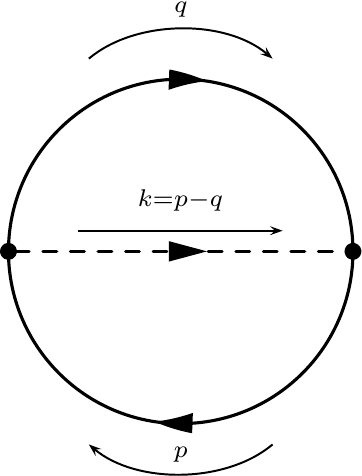}}
 ~=~-|y|^2\,
 \SumInt\limits_{\{p\}}\SumInt\limits_{\{q\}}
 \SumInt\limits_{k}\delta^{(4)}(p-q-k)\,
 \Tr\left[\frac{1}{\cancel{p}}\,P_\mathrm{L}\,\frac{1}{\cancel{q}}\,P_\mathrm{R}\,
 \frac{1}{k^2}\right]\;.\notag\\[-\dimexpr2\ht\strutbox\relax]
\end{align}
By convention~\cite{Laine:2016hma}, the momenta of the fermions are
indicated by curly brackets. The integrand of the 2{\tdash}loop integral
\eqref{eq:DeltaF1} can be rewritten as
\begin{align}
 P&~=~\Tr\left[\frac{1}{\cancel{p}}\,P_{\text{L}}\,
 \frac{1}{\cancel{q}}\,P_{\text{R}}\,\frac{1}{(p-q)^2}
 \right]
 ~=~\frac{\Tr\left[P_{\text{L}}\,\cancel{p}\,\cancel{q}\right]}{p^2\,q^2\,(p-q)^2}
 \notag\\
 &~=~\frac{2p\cdot q}{p^2\,q^2\,(p-q)^2}
 ~=~\frac{p^2+q^2-(p-q)^2}{p^2\,q^2\,(p-q)^2}
 ~=~\frac{1}{q^2\,k^2}+\frac{1}{p^2\,k^2}-\frac{1}{p^2\,q^2}\;,
\end{align}
where $k=p-q$ is the 4{\tdash}momentum of the scalar. So we are left with integrals of
the type
\begin{align}
 I~ \equiv &~\SumInt\limits_{\{p\}}\SumInt\limits_{k}\frac{1}{p^2\,k^2}
 +\SumInt\limits_{\{q\}}\SumInt\limits_{k}\frac{1}{q^2\,k^2}
 -\SumInt\limits_{\{p\}}\SumInt\limits_{\{q\}}\frac{1}{p^2\,q^2}
 \notag\\
 ~=&~2\,\left(\SumInt\limits_{\{p\}}\frac{1}{p^2}\right)\cdot\left(\SumInt\limits_{k}\frac{1}{k^2}
 \right)
 -
 \left(\SumInt\limits_{\{p\}}\frac{1}{p^2}\right)\cdot
 \left(\SumInt\limits_{\{q\}}\frac{1}{q^2}\right)
 \;.
\end{align}
To evaluate them, we use the  standard integrals in \cite{Laine:2016hma},
\begin{align}\label{eq:StandardLoopIntegralsTFT}
 I_T(0)&~=~\SumInt\limits_p\frac{1}{p^2}~=~\frac{T^2}{12}
 \quad\text{and}\quad
 \widetilde{I}_T(0)~=~\SumInt\limits_{\{p\}}\frac{1}{p^2}~=~
 -\frac{T^2}{24}
\end{align}
for bosons and fermions, respectively. This yields the contribution of one set
of chiral fermions with Yukawa coupling $y$ to the free energy as
\begin{align}
 \Delta \mathcal{F}&~=~\frac{5\,|y|^2}{576}\,T^4\;.
\end{align}
Hence, in the SM the contribution of a quark $q$ with Yukawa
coupling $y_q$ is given by\footnote{There is a
color factor of 3 and another factor 2 because the Higgs and one of the fermions are
$\text{SU}(2)_\mathrm{L}$ doublets.}
\begin{align}
 \Delta \mathcal{F}_\text{SM}&~\simeq~\frac{5\,|y_q|^2}{96}T^4\;.
\end{align}

\section{Analytic Solution for Flavon Oscillation} 
\label{sec:oscillations}

In the regime where the flavon decay can be neglected ($H \gg \Gamma_\sigma$)
and where the universe is radiation{\tdash}dominated ($H= T^2/M_H$), the equation of
motion \eqref{eq:sigmaEOM} can be solved exactly to find the oscillation of the
flavon about its zero{\tdash}temperature minimum. 
In this appendix we describe the asymptotic solutions for 
$\sigma(T)$ at high and low temperatures, and show how the amplitude of the flavon oscillations 
can be derived for a given set of initial conditions.

\begin{figure}
\centerline{\subfigure[~$\omega_{0,1,2}$.]{
\includegraphics[scale=0.6]{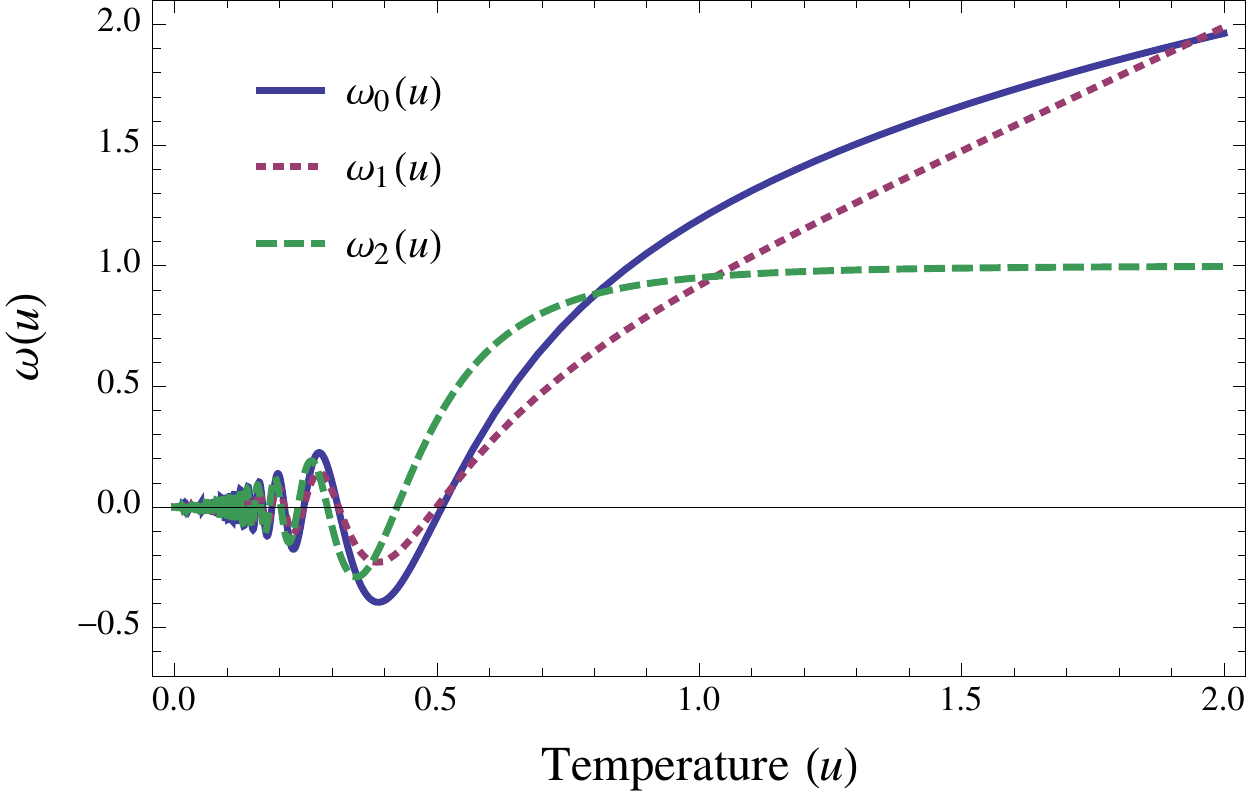} 
}\quad
\subfigure[~$\omega(z)$.]{
\includegraphics[scale=0.6]{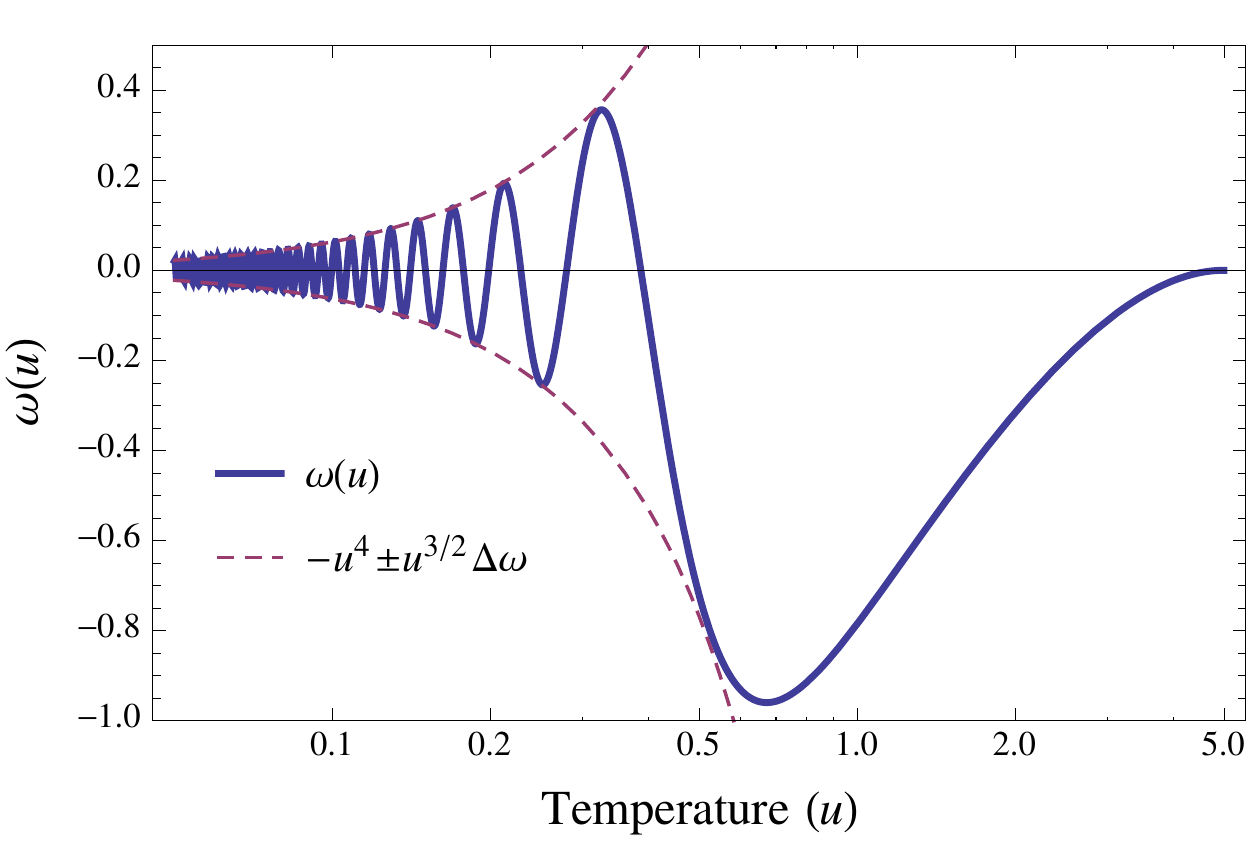} 
}}
\caption{In (a) we show the three functions $\omega_{0,1,2}$ which enter
into the general solution, $\omega=\omega_0 + c_1 \omega_1 + c_2\omega_2$.  In
(b) we take the initial condition $\omega(\z_0)=\omega'(\z_0)=0$ at
$\z_0=5.0$. The dashed lines represent the envelope $-\z^4\pm \z^{3/2}
\Delta\omega$, which accurately describes the low{\tdash}temperature ($\z<1$) oscillation about
the minimum of the finite{\tdash}temperature scalar potential.
}
\label{fig:analytic}
\end{figure}

We define the dimensionless variables $\omega$ and $z$ which replace $\sigma$
and $T$, such that all of the coefficients in the equation of motion are $+1$,
\begin{align}
\omega ~ \equiv ~\frac{ \Lambda}{\alpha M_H^2} \sigma\; ,&&
T_*~ \equiv ~ \sqrt{\ms M_H}\; , &&
\z ~ \equiv ~ \frac{T}{T_*}\; , &&
\frac{\dd^2 \omega}{\dd \z^2} + 
\frac{\omega}{\z^6} + \frac{1}{\z^2} ~= ~0\;. \label{eq:diffeq}
\end{align}
A general solution takes the form 
\begin{equation}
\omega(\z) ~=~ \omega_0(\z) + c_1\, \omega_1 (\z) + c_2\, \omega_2(\z)\;,
\end{equation}
where coefficients $c_1$ and $c_2$ are determined by the boundary conditions,
and where
\begin{align}
\omega_0(\z) ~=~&  
\frac{\pi}{\Gamma(1/4) \sqrt{\z}} J_{-\frac{1}{4} }\!\left(\frac{1}{2 \z^2} \right) \, {}_1\hspace{-0.04cm} F_2 \!\left( \frac{1}{4} ; \frac{5}{4}, \frac{5}{4} ; - \frac{1}{16 \z^4} \right) 
\notag\\
&{}+ \frac{\pi \sqrt{\z}}{8}  J_{\frac{1}{4} }\!\left(\frac{1}{2 \z^2} \right) \, G^{2,0}_{1,3}\! \left( \frac{1}{16 \z^4} \left| \begin{array}{c}  1 \\ 0,0, \frac{1}{4}    \end{array} \right. \right)
\;,
\label{eq:analyticW}\\
\omega_1(\z) ~=~&  \Gamma(3/4)\sqrt{ \frac{\z}{2}}  \,
J_{-\frac{1}{4}}\!\left(\frac{1}{2 \z^2} \right)\;, \qquad
\omega_2(\z) ~=~ \Gamma(5/4) \sqrt{ 2\z}  J_{\frac{1}{4}}\! \left(\frac{1}{2
\z^2} \right)\;  .
\end{align}
The function $\omega_0$ is expressed in terms of 
hypergeometric~$F$ and Meijer~$G$ functions. A plot of the functions $\omega_{0,1,2}$ is shown in Figure~\ref{fig:analytic}.

At high temperatures $\z\gg1$, the functions $\omega_i$ have asymptotic series expansions:
\begin{align}
\omega_0(\z) ~ \rightarrow ~& \ln\,\z +1.2774 - \frac{\ln\, \z}{20 \z^4} - \frac{0.0864}{
\z^4}  + \mathcal O(\z^{-8})\; , \\
\omega_1(\z) ~ \rightarrow ~& \z - \frac{1}{12 \z^3} + \mathcal O(\z^{-7})\; , \\
\omega_2(\z) ~ \rightarrow ~& 1 - \frac{1}{20 \z^4}  + \mathcal O(\z^{-8})\;.
\end{align}
These solutions correspond to the era in which the flavon mass is small compared to the Hubble rate, $H\gg \ms$, so that the $\ms^2 \sigma$ term in \eqref{eq:sigmaEOM} is negligible.

When the temperature drops below $\z\lesssim 1$, the flavon mass becomes relevant and the field begins to oscillate. In this limit the solutions $\omega_i$ are well approximated by
\begin{subequations}
\begin{align}
\omega_0(\z) ~\approx~ & 1.607 \z^{3/2} \cos\left(\frac{1}{2 \z^2} - \frac{
\pi}{8} \right)\;, \\
\omega_1(\z) ~\approx~ & 0.978 \z^{3/2} \cos\left(\frac{1}{2 \z^2} - \frac{
\pi}{8} \right)\;, \\
\omega_2(\z) ~\approx~ & 1.446 \z^{3/2} \cos\left(\frac{1}{2 \z^2} - \frac{3
\pi}{8} \right)\;, 
\end{align}
\end{subequations}
which satisfy the low{\tdash}temperature limit of \eqref{eq:diffeq}, $\z^2 \frac{\dd^2\omega}{\dd\z^2} + \omega/\z^4 =\mathcal O(\z^{3/2}) \ll1$. Conveniently, the phases of $\omega_0$ and $\omega_1$ approach each other as $\z\rightarrow 0$, so that the amplitude of the total oscillation can be written in terms of $c_1$ and $c_2$:
\begin{subequations}
\begin{align}
\omega(\z) &~=~ -\z^4+ \z^{3/2} \Delta\omega  \cos\left(\frac{1}{2 \z^2} -
\phi_0 \right)\;, \\
\Delta\omega^2 &~=~   (1.607 + 0.978 c_1)^2 + (1.446 c_2)^2  + 2.045 (1.607 +
0.978 c_1) c_2\;,
\label{eq:lowzsolution}
\end{align}
\end{subequations}
where the phase $\phi_0$ of the general solution is also determined by $c_1$ and $c_2$.
At temperatures $H\lesssim \ms$, the flavon oscillates about the minimum of its
finite{\tdash}temperature scalar potential, $\omega=-\z^4$, rather than $\omega=0$: this distinction becomes important for small reheating temperatures $T_R \ll \sqrt{\ms M_H}$.

Given specific boundary conditions $\omega(\z_0)$ and $\omega'(\z_0)$ at some
initial $\z_0 = T_R/ \sqrt{\ms M_H}$, the coefficients $c_1$ and $c_2$ can be
derived from the exact solutions $\omega_{i}$. For $\z_0 \gg 1$ it is easier to
use the asymptotic solution $\omega(\z) \simeq \ln\, \z + c_1 \z + c_2 + 1.277$
instead, with the result
\begin{align}
c_1 ~=~ \omega'(\z_0)  - \frac{1}{\z_0} \quad\text{and}\quad
c_2 ~=~ \omega(\z_0) - \z_0\, \omega'(\z_0) - \ln\, \z_0 - 0.277\; .
\label{eq:boundaryC}
\end{align}
As we discuss in Section~\ref{sec:scalarF}, a modest lower bound on the
amplitude of the oscillation can be found by assuming the initial condition
$\omega(\z_0) = \omega'(\z_0)=0$, which corresponds to the $\z\gg 1$ asymptotic
solution $\omega(\z) = \ln (\z/\z_0) - \z/\z_0 + 1$. The amplitude of the
oscillations when the temperature falls below $\z\lesssim1$ can be taken simply
from \eqref{eq:lowzsolution}. 
More general initial conditions such as $\omega'(\z_0)\neq 0$ typically produce
larger oscillations, though it is possible to fine{\tdash}tune the initial conditions such that it is suppressed.

\begin{figure}[t]
\centerline{\subfigure[~]{
\includegraphics[scale=0.6]{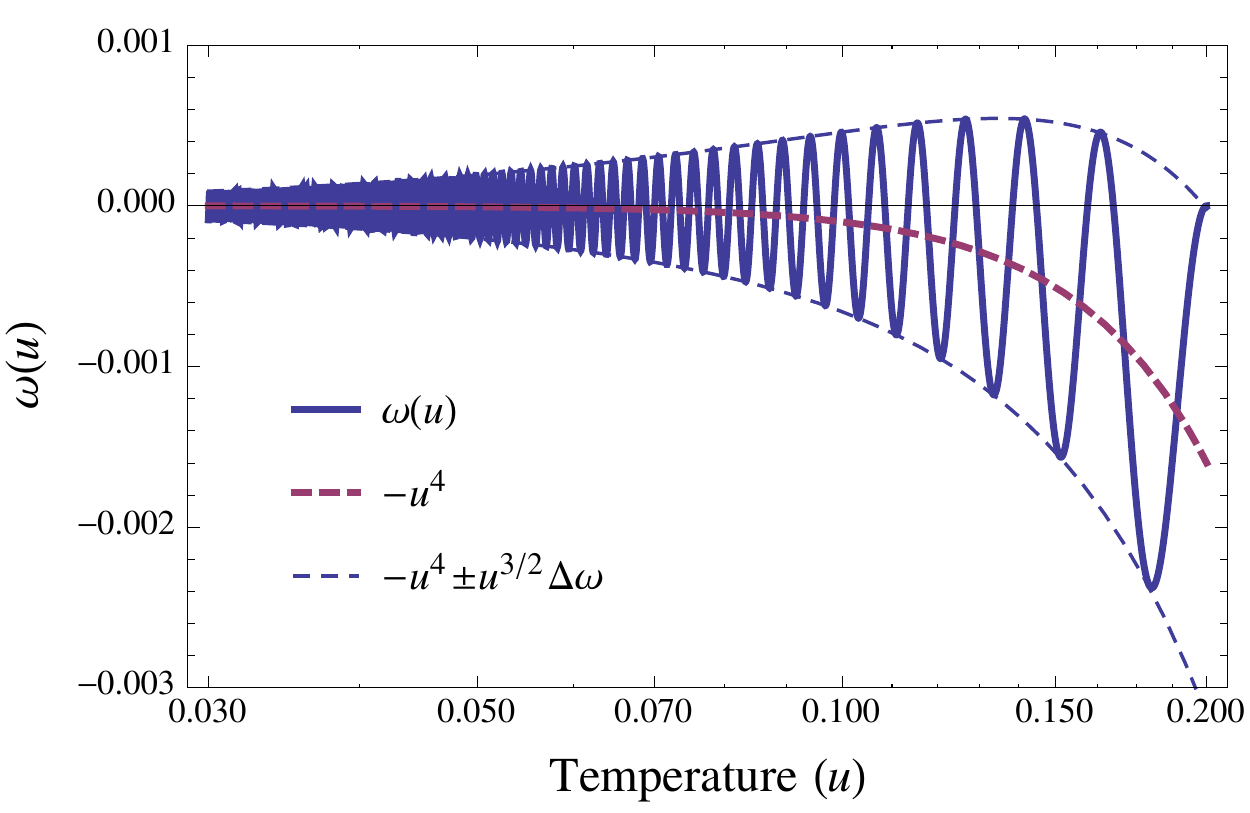} 
}\quad
\subfigure[~\label{fig:OscillationAmplitude}]{
\includegraphics[scale=0.6]{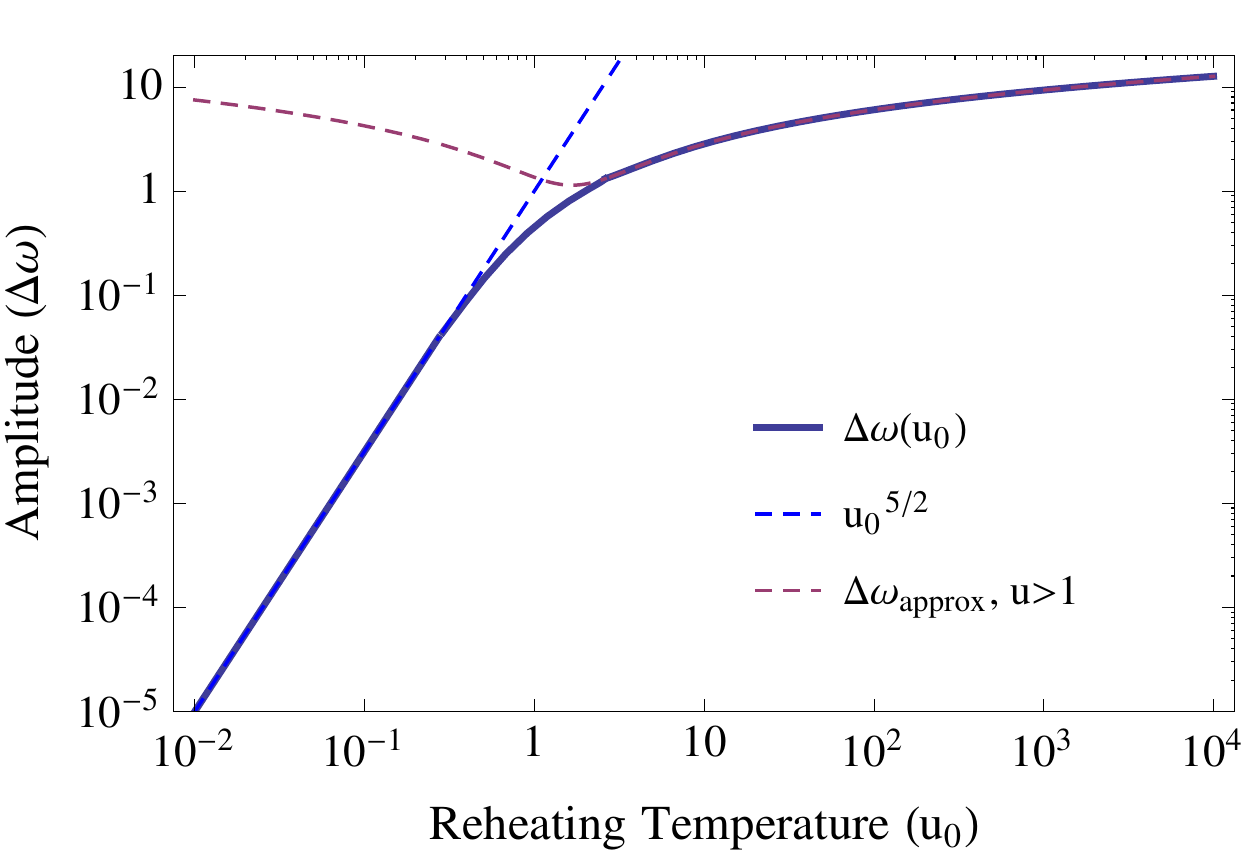} 
}}
\caption{In (a) we take $\omega(0.2)=\omega'(0.2)=0$, and show that for low reheating temperatures ($\z_0 \ll 1$) the flavon oscillates about $\omega=-\z^4$ with an amplitude proportional to $\z_0^{5/2}\z^{3/2}$.
In (b) we plot the amplitude $\Delta\omega$ of the flavon oscillations as a
function of the rescaled reheating temperature, $\z_0$, again using the
assumption that the flavon begins at rest at the minimum of its
zero-temperature scalar potential. The solid blue line utilizes the exact
solution to the equations of motion; the dotted lines show the low-temperature
and high-temperature approximations, respectively.
}
\label{fig:amplitude}
\end{figure}

If the reheating temperature is small to begin with, $\z_0 \ll 1$, the flavon oscillates freely about the minimum of its finite temperature potential, $\omega(\z)=-\z^4$. Taking the $\omega(\z_0) = \omega'(\z_0)=0$ initial condition, the amplitude of the flavon oscillations is $\z_0^{5/2} \z^{3/2}$.
The left panel of
\Cref{fig:amplitude} shows $\omega(\z)$ with $\z_0=0.2$, along with the envelope function $-\z^4 \pm \Delta \omega$ with $\Delta\omega = \z_0^{5/2}$.

In the intermediate range $0.4 \lesssim \z_0 \lesssim 2.0$, we find the flavon
amplitude $\Delta\omega$ by solving for $c_1$ and $c_2$ exactly. This result is
shown in the right panel of
\Cref{fig:amplitude} along with the $\z_0\ll1$ and $\z_0\gg1$ approximations to $\Delta \omega$.

As in freeze{\tdash}in flavon production, it is convenient to use the flavon yield
$Y_\sigma= n_\sigma /s$ to describe the flavon abundance. From
\eqref{eq:lowzsolution}, it can be seen that $Y_\sigma$ is constant with respect
to temperature,
\begin{equation}
Y_\sigma ~= ~\frac{\rho_\sigma}{\ms s}~ =~ \alpha^2 \frac{A_\star \mpl
\ms}{\Lambda^2} \left( \frac{M_H}{\ms} \right)^{3/2} \Delta\omega^2\;, 
\end{equation}
where 
\begin{equation}
\Delta\omega^2 ~=~  
\left\{ \begin{array}{l c l}  \left( T_R/T_* \right)^{5} && \text{for } T_R
\lesssim \frac{1}{2}\, T_*\;, \\ 
\Tstrut
2.09 \ln^2\left(T_R/T_*  \right) && \text{for } T_R \gtrsim 2 \, T_*\; .
\end{array} \right. 
\end{equation}

\section{Freeze-In With Correct Statistical Factors}
\label{sec:statistics}

In thermal equilibrium, the phase space density for fermions or bosons is given
by $f_{\psi,\phi}(E) = (\mathrm{e}^{E/T} \pm 1)^{-1}$,
with $E = \sqrt{p^2 + m^2}$ for massive particles. In the limit $T\ll m$, $f_{\psi,\phi}(E)$ both approach the Maxwell--Boltzmann (MB) distribution, $f_{\psi,\phi}(E) \approx \mathrm{e}^{-E/T}$, which simplifies the Boltzmann equation significantly.
For freeze{\tdash}in processes driven by nonrenormalizable operators, however, the
freeze{\tdash}in yield is typically driven by high{\tdash}temperature production, $T\sim T_R$, which in our case of weakly coupled flavons implies the relativistic limit, $T\gg \ms$.

In this section, we use correct fermionic/bosonic statistics to
calculate the contribution to the freeze{\tdash}in yield from relativistic $q \bar q
\rightarrow \sigma \Phi$ scattering, in order to estimate the error introduced by approximating them as Maxwell--Boltzmann.
With its fermionic initial state and
bosonic final state, this process receives the most significant modification from
the statistical factors, and can therefore be used to infer an upper bound on
the total freeze{\tdash}in yield. Conveniently, the squared amplitude for this process is independent of the angular coordinates:
\begin{equation}
\abs{ \mathcal M\left(q_i\bar q_j\rightarrow \sigma \Phi^{(*)}\right) }^2 ~=~ \frac{24
\abs{g_{ij}^{u/d}}^2 s}{\Lambda^2}\;,
\end{equation} 
where $s=(p_A+p_B)^2 = (p_1+p_2)^2$. Here $p_A$ and $p_B$ refer to the momenta of the incoming quark and antiquark, while $p_1$ and $p_2$ refer to the flavon and Higgs doublet, respectively.
The evolution of the flavon number density $n_\sigma$ is given by the Boltzmann equation,
\begin{align}
\dot n_\sigma + 3 H n_\sigma ~=~& \int\!  \dd \Pi_A\, \dd \Pi_B\, 
\dd \Pi_1\, \dd \Pi_2\, f_A(E_A)\, f_B (E_B)\, \left[ 1 + f_2 (E_2) \right] 
\nonumber\\&\; \times
(2\pi)^4 \delta^{(4)} (p_A + p_B - p_1 - p_2)\, \msq\;,
\end{align}
where $\dd \Pi_i = \frac{\dd^3 p_i}{(2\pi)^3 2 E_i}$ is the Lorentz{\tdash}invariant
phase space. 
Defining the variables $E_+ = E_A + E_B$ and $E_- = E_A - E_B$, the initial
state $\dd^3 p_A\, \dd^3 p_B$ integral simplifies to
\begin{align}
\dd \Pi_A\, \dd \Pi_B ~=~ \frac{  \dd E_+\, \dd E_-\, \dd s}{2^7 \pi^4 }\;.
\end{align}
We also define $\dd^3 p_2 = (2\pi) E_2^2 \dd E_2 \dd\!\cos\theta_2$, where
$\theta_2$ is the angle between $\vec{p}_2$ and the center{\tdash}of{\tdash}mass 3{\tdash}momentum $\vec{p}_1+\vec{p}_2$, so that the remaining $\delta(E_+ - E_1 - E_2)$ can be used to perform the angular integration.
In terms of the dimensionless variables $z=\sqrt{s}/T$, $y=E_+/T$, $w=E_-/T$, and $\tau=E_2/T$, the Boltzmann equation is written:
\begin{align}
\frac{\dot n_\sigma + 3 H n_\sigma}{T^4}\; =\; \frac{1}{2^9 \pi^5}
\int\limits_0^\infty \! \dd z\,z\, \msq \int\limits_z^\infty\! \frac{\dd y}{\sqrt{y^2 - z^2}} 
\int\limits_{\mathclap{-\sqrt{y^2 - z^2}}}^{\mathclap{\sqrt{y^2 - z^2}}}\! \frac{\dd w}{\mathrm{e}^y + 1 + 2 \mathrm{e}^{y/2} \cosh \frac{w}{2} } 
\int\limits_{\tau_\text{min}}^{\tau_\text{max}} \! \frac{\dd \tau}{1 - \mathrm{e}^{-\tau}}
\;,
\label{eq:boltz2}
\end{align}
where
\begin{equation}
\tau_{\text{max,min}} = \frac{y\pm \sqrt{y^2 - z^2}}{2}\;.
\end{equation}
Both the $u$ and $\tau$ integrals can be completed analytically, with the result
\begin{align}
\frac{\dot n_\sigma + 3 H n_\sigma}{T^4} \;=\;& \frac{1}{2^6 \pi^5} 
\int\limits_0^\infty\! \dd z\,z\, \msq \int\limits_z^\infty\! \frac{\dd y}{\sqrt{y^2 - z^2}} 
\nonumber\\
&{} 
\frac{\arctanh \left( \tanh\frac{y}{4} \tanh\frac{\sqrt{y^2 - z^2}}{4}  \right) }{\mathrm{e}^y - 1}
\ln\left( \frac{\exp\frac{y + \sqrt{y^2 - z^2}}{2}  - 1}{\exp\frac{y - \sqrt{y^2 - z^2}}{2}  - 1} \right)
\;.
\label{eq:boltzZY}
\end{align}
Noting that $\msq\sim z^2 T^2/\Lambda^2$, the above integral can be integrated numerically:
\begin{equation}
\frac{\dot n_\sigma + 3 H n_\sigma}{T^4} = \left( 7.8 \times 10^{-4} \right) \times \frac{24 g_{ij}^2 T^2}{\Lambda^2}
\;.
\label{eq:bzfinal}
\end{equation}
When the MB approximation is taken in \Cref{eq:boltz2}, the result is smaller,
\begin{align}
\frac{\dot n_\sigma + 3 H n_\sigma}{T^4} &\;\approx\; \frac{1}{2^9 \pi^5} 
\int\limits_0^\infty\! \dd z\,z\, \msq 
\int\limits_z^\infty\! \frac{\dd y}{\sqrt{y^2 - z^2}} 
\int\limits_{\mathclap{-\sqrt{y^2 - z^2}}}^{\mathclap{\sqrt{y^2 - z^2}}}\! \frac{\dd w}{\mathrm{e}^y } 
\int\limits_{\tau_\text{min}}^{\tau_\text{max}}\! \dd \tau \\
&~\approx~ \frac{1}{2^8 \pi^5} 
\int\limits_0^\infty\! \dd z\,z\, \msq z K_1(z)
~=~ \left(2.04 \times 10^{-4} \right) \times \frac{24 g_{ij}^2 T^2}{\Lambda^2} \;.
\label{eq:Mboltz2}
\end{align}
Our use of Maxwell--Boltzmann statistics in \Cref{sec:yUV} causes us to
 underestimate the $Q_L \bar q \rightarrow \sigma \Phi$ scattering rate by a
 factor of $3.8$.

Of the six processes identified in \Cref{eq:UVprocess}, only the two with purely
bosonic final states receive this factor of $\sim$four enhancement.  To
calculate the contributions from the remaining processes one must additionally
perform the angular integral, which becomes nontrivial when the initial and
final states both include a fermion. Considering that the total flavon yield is
dominated by $Y^\text{osc}$ rather than $Y^\text{fi}$ in most of the parameter
space, we leave this more precise determination of the relativistic freeze{\tdash}in
yield to future studies.

\bibliographystyle{h-physrev} 
\bibliography{flavon_BBN}

\end{document}